\documentclass[10pt]{iopart}
\usepackage{graphics}
\usepackage{graphicx}
\usepackage{hyperref}
\usepackage{amssymb}
\usepackage[numbers,square,comma,sort&compress]{natbib}
\usepackage{epstopdf}

\begin{document}
\title[The Physics of the Colloidal Glass Transition]{The Physics of the Colloidal Glass Transition}
\author{Gary L. Hunter and Eric R. Weeks}
\address{Department of Physics, Emory University,
Math \& Science Center
400 Dowman Dr., N201
Atlanta, GA 30322}
\ead{weeks@physics.emory.edu}
\begin{abstract}

As one increases the concentration of a colloidal suspension, the
system exhibits a dramatic increase in viscosity.  Structurally,
the system resembles a liquid, yet motions within the
suspension are slow enough that it can be considered essentially
frozen.  This kinetic arrest is the colloidal glass transition.
For several decades, colloids have served as a valuable model
system for understanding the glass transition in molecular systems.
The spatial and temporal scales involved allow these systems
to be studied by a wide variety of experimental techniques.
The focus of this review is the current state of understanding of
the colloidal glass transition.  A brief introduction is given to important
experimental techniques used to study the glass transition in
colloids.  We describe features of colloidal systems near and
in glassy states, including tremendous increases in viscosity and
relaxation times, dynamical heterogeneity, and ageing, among others.
We also compare and contrast the glass transition in colloids to that in molecular liquids.
Other glassy systems are briefly discussed, as well as recently
developed synthesis techniques that will keep these systems rich
with interesting physics for years to come.

\end{abstract}

\submitto{\RPP}
\maketitle

\tableofcontents
\newpage

\section{Introduction}\label{intro}

\subsection{What is the Colloidal Glass Transition?}

Imagine you have a bucket of ink.  Ink is composed of colourful
micron-sized particles in water.  If you let the water evaporate
from the ink, the ink becomes more and more viscous and at some point,
it is still damp but no longer flows easily.  This increase in
viscosity as the water is removed is the colloidal glass transition,
and in many respects is analogous to how window glass solidifies
as it is cooled from a high temperature.

A colloidal suspension is composed of small solid particles in
a liquid, like ink or paint.  The key control parameter is the volume fraction $\phi$:  the fraction of volume occupied
by the solid particles.  Samples with a larger volume fraction will
have a larger viscosity, and this viscosity grows quite dramatically
as $\phi \rightarrow \phi_g \approx 0.58$.  As the glass transition
volume fraction $\phi_g$ is approached, the sample's behaviour
parallels the glass transition of more traditional (molecular or polymer)
glass-forming systems \cite{sciortino05}.  In a chunk of window
glass, the atoms are arranged in an amorphous fashion; likewise,
in a dollop of glassy colloidal paste, the colloidal particles are
arranged in an amorphous way.  Given the size of colloidal particles
($\sim 10~\rm{nm} - 10~\mu$m diameter), they can be studied using
a variety of techniques that are difficult or impossible to adapt
to molecular glass-formers.

In the following subsections, we introduce basic concepts such as
colloids, glasses, and some relevant physics, before proceeding with
the rest of the review.

\subsection{Introduction to the Glass Transition}\label{glasstransition}


Upon slow cooling or compression, many liquids freeze -- that is,
the molecules constituting the liquid rearrange to form an ordered
crystalline structure.  In general, nucleating a crystal requires
undercooling.  Some materials can be substantially undercooled
without crystal nucleation; alternatively, a sample can be cooled faster that nucleation can occur.
In such situations, the liquid is termed {\it supercooled}.  If the
sample is sufficiently cold and cooling is adequately rapid, the
material can form a glass: the liquid-like structure is retained
but the microscopic dynamics all but cease.  This sudden arrest
is the {\it glass transition}, and the temperature at which
it occurs is the glass transition temperature, $T_g$.  As the
liquid is cooled toward $T_g$, its viscosity rises smoothly and
rapidly, and below $T_g$ the sample's viscosity becomes so high
that for most practical purposes it is considered a solid.
The science of the glass transition is discussed in many
review articles \cite{gotze92,angell95,stillinger95,ediger96,
angell00q,debenedetti01,andersen05,dyre06,berthier11rmp}.  {\it Supercooled liquid}
refers to a system under conditions for which it still
flows, but for which the liquid is a metastable state and the
thermodynamically preferred state is a crystal.  The study of the
glass transition then is the study of how a supercooled liquid
changes as the temperature $T$ is decreased toward $T_g$, and
the study of glasses is the study of materials under conditions
where $T<T_g$.  Glasses can also be formed at constant $T$ by
increasing pressure \cite{roland05,win06}.

Calling a glassy material a ``solid'' depends on time scales, and perhaps
one's patience~\cite{lindsay82}.  Window glass, a vitreous
form of silicon dioxide, is of course the quintessential example
of glass.  It is sometimes claimed that very old windows are thicker at the bottom due to flow of glass.  However,
the thickness variations in antique windows are the result of
a particular manufacturing method rather than the result of the
glass flowing over long times~\cite{zanottoglass1,zanottoglass2}.
A more instructive example of glassy behaviour and time scales
can be seen in pitch, a bituminous tar.  Like window glass, pitch
is unmistakably solid to the touch -- if struck with a hammer,
it will shatter.  However, for over 80 years a funnel filled
with pitch has been dripping at a rate of roughly one drop every
100 months, yielding a very approximate viscosity of $10^{11}$
times that of water.  The so-called ``Pitch Drop Experiment"
has been housed at the University of Queensland in Brisbane,
Australia since 1927 \cite{pitchdrop}.

\subsection{Introduction to Colloids}\label{definecolloid}

The term {\it colloid} describes a wide range of multiphasic
substances composed of particles (solid, liquid or gaseous)
roughly 10 nm$-10$~$\mu$m in size dispersed in a continuous phase.
Depending on the state of matter of the various phases, colloids can
be divided into several categories, including, but not limited to:
\begin{list}{$\bullet$}{\setlength{\itemsep}{0ex}}
\item Suspensions/Dispersions -- Solid particles in a liquid
(this review's main focus)
\item Emulsions -- Liquid droplets in an immiscible liquid 
\item Foams -- Gas bubbles in a liquid or solid medium 
\item Aerosol -- Liquid droplets or solid particulates in a gas 
\end{list}
Hence, {\it colloid} is equally apt to refer to a variety of
systems: such as ink, paint, peanut butter, milk,
blood (suspensions); Styrofoam\texttrademark, shaving cream and ice
cream (foams); mayonnaise and hand lotion (emulsions); hair spray
and smoke (aerosols).  For this range of size, colloids behave
as systems of ``classical'' particles where quantum mechanical
effects can be largely ignored, though it is important to understand
the role of quantum phenomena such as van der Waals attractions.
More importantly, colloidal particles are small enough that thermal
fluctuations are extremely relevant.  For example in a suspension,
random collisions between solid particles and solvent molecules
lead to Brownian motion, easily observed in experiments.

Aside from the everyday items mentioned previously,
industrial processes such as liquid and mineral purification, oil
recovery and processing, detergency, and even road surfacing employ
colloids to varying degrees~\cite{cosgrovebook}.  Dense colloidal
suspensions can be heated and allowed to flow while retaining
some rigidity.  Hence, they can be moulded, extruded, and 
subsequently solidified to form a multitude of components.
The manufacture of many types of optics, insulators, bricks, and
ceramics involve colloids~\cite{larson98}.

While these examples span a wide range of useful materials, colloids also find use in laboratories as models
for phases of matter.  During the late 1960's and early 1970's,
experiments demonstrated that structures
in colloidal suspensions can be analogous to those in atomic
systems~\cite{krieger1969,krieger1971,kose1973jcis},
leading to extensive use of colloids over the
next decade as model liquids and crystals~\cite{hastingsmacroions,ackerson1986,vanwinkle1986,murray1987,chaikin1989,murray1990,marshall1990,murray1996arpc,pusey86}.
In 1982, Lindsay and Chaikin combined two different sizes of
charged colloidal particles and observed a glassy phase (amorphous
structure, finite rigidity)~\cite{lindsay82} in agreement with
subsequent simulations~\cite{rosenberg89}.  Later in 1986 and
1987, experiments by Pusey and van Megen demonstrated
a hard-sphere colloidal glass transition in a concentrated sample
of uncharged colloids~\cite{pusey86,pusey1987,pusey1987jphysique}.


\subsection{Basic Physics:  Hard-Sphere-like Colloids}\label{colloidsasmodel}

Perhaps the simplest interaction between two
particles is that of {\it hard-spheres}~\cite{bernal64}.
If $r$ defines the distance between two sphere centres, and $\sigma$
is the sum of the two sphere radii, the hard-sphere potential is
given by
\begin{equation}
V(r) = \left\{ \begin{array}{rl}
 \infty , &\mbox{ if $r \leq \sigma$} \\
 0 ,&\mbox{ otherwise,}
       \end{array} \right.
\end{equation}
which is to say that the only restriction placed upon the system
is that particles cannot interpenetrate.  Hence, all allowable
configurations have identically zero potential energy.  From a
viewpoint of statistical mechanics, this implies that the free
energy, $F = U-TS = 3Nk_B T - T S = (const - S)T$, is governed entirely by entropy~\cite{lowenhardspheres,donev07}, which
for monodisperse systems (systems of a single particle size)
means that the only control parameter is volume fraction
\cite{wood1957,alder1957,hoover67,sciortino05}.  Volume
fraction, $\phi = NV_p/V$, is a dimensionless analogue of
particle number density, where $N$ is the number of particles
in the system, $V_p$ is the single particle volume, and $V$
is the total system volume.  (Note that the particle size
controls how fast a system evolves due to diffusion, but does
not control the phase behaviour; see section~\ref{basicphysics}
for discussion of particle size effects.)  

\begin{figure}
\centerline{
\includegraphics[width=.9\textwidth]{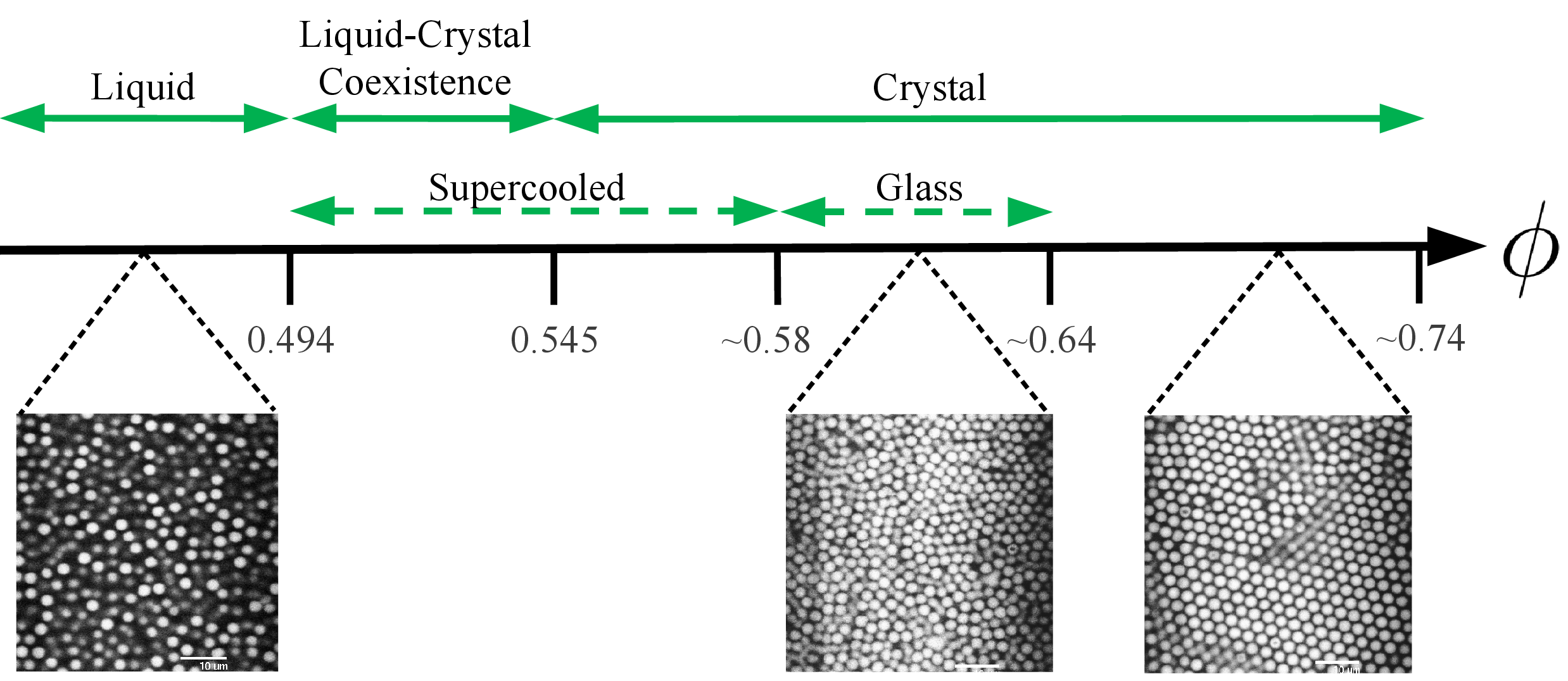}}
\caption{ (Colour online) Top:  Phase diagram of monodisperse
hard-spheres as a function of volume fraction, $\phi$.
Solid arrows indicate equilibrium states, whereas dashed arrows
are non-equilibrium states.  Note that the existence of the
glassy state requires some polydispersity (perhaps at least
8\%); a more monodisperse sample will eventually crystallize
\cite{bolhuis96,auer01b,schope07,zaccarelli09,pusey09}.  However,
polydispersity also shifts the boundaries between liquid and crystal
to slightly higher values \cite{bolhuis96,fasolo03,sollich10}.
On the other hand, adding a slight charge to the particles shifts
the phase boundaries to lower values
\cite{hynninen03,hernandez09}.
Bottom: Confocal micrographs of the analogous hard-sphere phases
in a colloidal suspension with 5\% polydispersity.
}
\label{hardspherephasediagram}
\end{figure}

The phase diagram for hard-spheres is shown in
figure~\ref{hardspherephasediagram} as a function of $\phi$.  Below the freezing point,
$\phi_{\rm freeze} = 0.494$, the suspension is a liquid.  Forcing the system
into a supercooled or glassy state requires increasing $\phi$ fast
enough to avoid crystallization.  The supercooled region persists
between $0.494 \leq \phi < \phi_g \approx 0.58$, whereas the glassy
region lies between $ \phi_g < \phi < \phi_{rcp} \approx 0.64$.  
The existence of a glassy phase for hard-spheres requires that the
sample be somewhat polydisperse, that is, the spheres must have a
distribution of sizes~\cite{bolhuis96,auer01b,schope07,zaccarelli09,pusey09}.
The upper bound of the glassy region is the volume fraction at random
close packing, $\phi_{rcp}$, the maximum density of a completely
random sphere packing~\cite{bernal60,torquato00,radin08}; the precise value of
$\phi_{rcp}$ depends
on the polydispersity~\cite{hermes10}.
Above $\phi_{rcp}$, samples must have domains of crystalline
structure, or, preferably from the thermodynamic point of view,
the sample may be entirely crystallized.  Density can be further
increased up the limit of hexagonal close packing, $\phi_{hcp}
= \pi/3 \sqrt{2} \approx 0.74$.  Hard-spheres are often simulated to study the glass transition~\cite{snook91,speedy98,doliwa98,lowen99,foffi04,xu09,zaccarelli09,pusey09}.

In many cases, colloidal
particles can be considered to be simple hard-spheres~\cite{pusey86,pusey1987,bryanthardsphere,underwood94colloids}.
The first experimental demonstration of a colloidal hard-sphere
glass transition was by Pusey and van Megen in the mid-1980's,
who essentially replicated the hard-sphere phase diagram using
colloidal samples~\cite{pusey86,pusey1987}; see for example the
pictures in figure~\ref{hardspherephasediagram}.  The system used in
these studies is particularly important for the following reasons:
the interactions between particles are of a simple, well-described
nature; the simplicity of the interaction allows for comparison to
a wide range of systems, and easy simulation with computers; they
can be studied by techniques such as microscopy, light scattering
and rheology -- that is, a single sample can be divided and studied
by an array of methods.  For these reasons, along with the fact
that particles are commercially available or can be synthesized
readily~\cite{antlsynthesis,bosma02,campbell02,elsesser10}, the
same types of colloids are still widely used today.

The particles used by Pusey and van Megen were composed of
poly(methyl methacryalate) (PMMA) and were sterically stabilized
by the addition of a thin surface layer ($\approx 10$~nm) of
poly-(12 hydroxystearic acid) (PHSA) to minimize aggregation due to
van der Waals forces.  It is this steric stabilization layer that
allows particles to be considered as hard-spheres, at least until they are forced close enough compress the PHSA~\cite{poon12}.  These colloids are stable in organic solvents
and can be somewhat tailored for experiments, such as being dyed for use in fluorescence microscopy~\cite{campbell02,elsesser10,dinsmore01}.

Solvent choice also allows for a greater degree of control.  Miscible solvents can be mixed to closely match the density of the particles, minimizing
gravitational effects that can be quite significant in studying
colloidal glasses~\cite{kegel04,zhu97,kegel00lang} (discussed below).  Solvents can also be blended to closely
match the refractive index of the particles, which both lessens
van der Waals attractions and allows for use in microscopy or
light scattering.

The size range and time scales that accompany colloidal particles are accessible to a variety of experimental techniques such as optical
microscopy or light scattering.  For example, a micron-sized
particle in water will diffuse its diameter in about a second,
which is easily observable for modern microscopes.  

It is important to note that colloidal systems
differ from their atomic counterparts in several ways~\cite{bartsch98cocis,schweizer07}.  First, short time motion
is diffusive in colloids, rather than ballistic.  Second,
hydrodynamic effects couple particle motions in complex ways~\cite{zaccarelli2009jcp}.  Simulations suggest that these two
differences are unimportant for studying the glass transition~\cite{binder98,szamel04,tokuyama03,tokuyama07,berthier07a,franosch08}
(see also discussion in section~\ref{simulations}).  A third difference
is that colloidal particles are most typically spherically
symmetric, and so the geometry of a molecule is usually not
replicated in the colloid (see section~\ref{otherglasses}
for recent exceptions).  Again, for many cases of interest,
this difference is immaterial when studying long-time dynamics;
certainly many glass transition simulations study particles with
spherically symmetric potentials.  A fourth difference is that
colloidal suspensions are always slightly polydisperse. This shifts the phase
transitions shown in figure~\ref{hardspherephasediagram} to
higher values of $\phi$~\cite{bolhuis96,fasolo03,sollich10},
and also in general frustrates crystallization~\cite{zaccarelli09,pusey09,henderson96,henderson98,auer01b}.
While this is a distinction in comparison to simple molecular
glass-formers, it is less of a distinction with simulations,
which often purposefully add polydispersity to frustrate
crystallization~\cite{kob95a,kob95b}.  Indeed, as noted in the caption of figure~\ref{hardspherephasediagram}, polydispersity
appears necessary for a hard-sphere glass transition; 
monodisperse samples always eventually crystallize~\cite{zaccarelli09,pusey09}.

A final distinction is
that colloidal samples are influenced by gravity.  As observed
in one experiment, a sample that was a colloidal glass on
Earth spontaneously crystallized in microgravity~\cite{zhu97}. Precisely matching the density of particles and solvent also potentially leads to crystallization and can have a striking influence on the ageing of a glassy colloidal sample~\cite{kegel04,kegel00lang} (see section~\ref{ageing} for discussion of ageing).  However, the interpretation of these
results is controversial.  The crystallization seen may be due to differing polydispersity which strongly influences nucleation~\cite{zaccarelli09} and may be a confounding variable in these experiments~\cite{pusey09,henderson98,auer01b,meller92,schope07}.
It may also be due to heterogeneous nucleation at the walls of the sample chambers~\cite{pusey86}.  However, given the robust similarities
between colloidal experiments and gravity-free simulations (described in detail in section~\ref{simulations}), it seems plausible
that gravity is typically not a critical factor, but we note this is debatable.

Although PMMA colloids are a widely used model
system, they are by no means the only colloidal system used to study
glass transition.  Other non-hard-sphere systems will be discussed
throughout this review, particularly in section~\ref{otherglasses}.

\subsection{More Basic Physics:  Diffusion and Sedimentation}
\label{basicphysics}

Two key concepts for thinking about the
colloidal glass transition are diffusion and sedimentation.
Diffusion sets the rate of the dynamics, and sedimentation can
limit the duration of experiments.

The size of colloidal particles is
such that they execute Brownian motion due to frequent, random
collisions with solvent molecules.  Because collisions are random
in magnitude and orientation, the average particle displacement
in a particular direction $\langle \Delta x \rangle $ is zero.
Instead, motion is often quantified by
the mean square displacement (MSD),
\begin{equation}\label{eqnmsd}
\langle \Delta x^2 \rangle = \langle [x(t + \Delta t) - x(t)] ^ 2 \rangle = 2D\Delta t.
\end{equation}

\noindent The angle brackets $\langle \rangle$ indicate an average
over all particles and all initial times $t$ for a particular
lag time $\Delta t$, and $D$ is the diffusion coefficient.  In three
dimensions, (\ref{eqnmsd}) becomes
\begin{equation}\label{msd3d}
\langle \Delta r^2 \rangle = 6D\Delta t.
\end{equation}

For a single particle of radius
$a$ immersed in a solvent of viscosity $\eta$, the diffusion
coefficient $D$ is given by the Stokes-Einstein-Sutherland
equation,
\begin{equation}
\label{stokeseinstein} 
D = \displaystyle \frac{k_B T}{6 \pi \eta a},
\end{equation}

\noindent where $k_B$ is Boltzmann's constant and $T$ is the
system temperature \cite{sutherland1905,einstein1905a}.  This
equation shows that $T$, $\eta$, and $a$ do not play a direct
role in the colloidal glass transition;  they only influence $D$,
which in turn sets a  time scale for particle motion.  This
time scale is the diffusive (or Brownian) time,

\begin{equation}
\tau_D = \displaystyle \frac{a^2}{6 D} = \displaystyle \frac{\pi
\eta a^3}{k_B T},
\label{taud}
\end{equation}

\noindent which is the average time needed for a particle to
diffuse its own radius [using $\langle \Delta r^2 \rangle = a^2$ in
(\ref{msd3d})].

\begin{figure}
\centerline{
\includegraphics[scale=0.4]{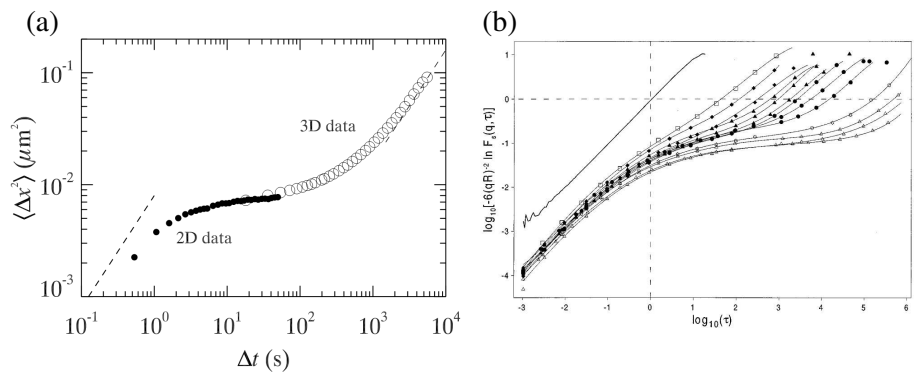}} \caption{(a) An example mean square
displacement from a sample with $\phi=0.52$, observed with confocal
microscopy.  Particles have radius $a=1.18$~$\mu$m; data
are from~\cite{weeks02sub}.  The 2D data (solid circles) is collected
at a fixed depth within the three-dimensional sample, while 3D data (open circles) is
collected over a fixed sample volume.  The 2D data can be acquired
more rapidly and probe shorter time scales, which
is why the two data sets extend over different ranges in $\Delta
t$.  The graph shows only the $x-$component of the MSD, but the
$y-$ and $z-$components are similar.  The dashed lines indicate a slope of 1.  
(b) Mean square displacements measured via light scattering.
Time scales are normalized by the diffusive time ($\tau_D$ = 0.0215 s) and
measured at $qR = 1.3$.  Volume fraction increases from left to
right: $\phi \approx 0$ (solid line), 0.466 (squares), 0.502,
0.519 (closed diamonds), 0.534, 0.538, 0.543, 0.548 (closed
triangles), 0.553, 0.558 (closed circles), 0.566 (stars), 0.573,
0.578, 0.583 (open triangles).  Figure (b) is reprinted with permission
from~\cite{vanmegen98}. Copyright 1998 by the  American Physical Society.
}
\label{examplemsd}
\end{figure}

For purely diffusive motion, such as in a dilute suspension,
the MSD scales with $\Delta t$.  Thus, on a log-log plot of $\langle
\Delta r^2 \rangle$ vs.~$\Delta t$, one expects a straight line with
a slope of unity.  Shown in figure~\ref{examplemsd}(a) is the MSD for
a colloidal sample at $\phi = 0.52$. At the smallest $\Delta t$,
the MSD shows diffusive behaviour, indicated by the dashed lines.
Note that the diffusion constant obtained from this short time-scale
motion, $D_S$, differs from that of [\ref{stokeseinstein}] for
$\phi>0$ due to hydrodynamic interactions between the particles~\cite{pusey83,snook83,beenakker83b,beenakker83,vanmegen85}.  $D_S$
drops to approximately 50\% of the value from [\ref{stokeseinstein}]
by around $\phi \approx 0.3$.  

As the lag time increases, a plateau develops in the data of
figure~\ref{examplemsd}(a) which
is indicative of particles being trapped in cages formed by
their neighbours.  At these time scales, particles are localized
and large cumulative motions are suppressed~\cite{berne66,
sjogren80,wahnstrom82,gotze92,rabani97,verberg99,schweizer03}.
At sufficiently long $\Delta t$,
particle rearrangements do occur, and so the MSD again increases,
eventually recovering diffusive behaviour.  Figure~\ref{examplemsd}(b) shows the MSDs measured with light scattering for samples at several volume fractions (see section~\ref{dls} for discussion of light scattering).  As $\phi$ increases from left to right, one observes a lengthening of the plateau and thus increasingly slowed dynamics.


The overall shapes of the MSDs in figure~\ref{examplemsd} are typical of dense suspensions.  In these cases,
one often characterizes the system in terms of a long time diffusion constant, defined as

\begin{equation}
D_L \equiv \lim_{t\to\infty} \frac{\langle \Delta r^2 \rangle }{6\Delta t}.
\label{dlong}
\end{equation}
This describes the motions within a system at times after the plateau in the MSD, as shown at large lag times in figure~\ref{examplemsd}.  The definition is especially useful when comparing  the relative importance of
externally applied motions to motions in a quiescent sample (see section~\ref{shear}).

If the particle size is doubled in a colloidal sample while keeping
the volume fraction $\phi$ constant, then the motion slows by
a factor of 2 on an absolute scale [from (\ref{msd3d}) and
(\ref{stokeseinstein})] and a factor of 8 relative to the particle
size [from (\ref{taud})].  However the overall appearance of the dynamics
(liquid-like or glassy) remains the same:  more specifically,
the behaviour of $\langle \Delta r^2 \rangle / a^2$ as a function of
$\Delta t / \tau_D$ is unchanged.  This suggests a useful
experimental technique:  to effectively explore long-time
dynamics, one might use smaller colloidal particles which diffuse
faster and reach the long-time behaviour on relatively short
experimental time scales.  In contrast, video microscopy
techniques (section~\ref{microscopy}) work better with slower
moving particles, so one typically uses larger particles in such
experiments.

The other important consideration for colloidal glass experiments
is sedimentation.  It is tricky to match the density of the
solvent to the density of the colloidal particles, and so over
time particles sink to the bottom of a sample chamber (or
float to the top).  This changes the local volume fraction, the
key control parameter, and so sedimentation is
important to understand for experiments.

The length scale over which gravity is important is set by
balancing the gravitational potential energy $\Delta \rho V g z$
with the thermal energy $k_B T$, where $\Delta \rho$ is the
density difference between the particle and the solvent, $V =
\frac{4}{3} \pi a^3$ is the volume of the particle, $g$ is the
acceleration of gravity, and $z$ is a height.  Solving this for
$z$ gives the scale height

\begin{equation}
z_0 = \displaystyle \frac{3}{4 \pi}
\displaystyle \frac{k_B T}{\Delta \rho a^3 g}.
\label{scaleheight}
\end{equation}

\noindent In equilibrium, $\phi$ varies over distances $\sim z_0$.  In particular, one expects to find
$\phi(z) \approx \phi_0 \exp(-z / z_0)$.  If a sample chamber has
a height much less than $z_0$, then sedimentation can probably
be ignored.  This is achieved using thin sample chambers, $\sim 200$~$\mu$m thick
typically.  Alternatively, one can use small particles;  as (\ref{scaleheight}) shows, $z_0 \sim a^{-3}$.  More careful matching of the
density of the solvent can minimize $\Delta \rho$; here, the chief
problem is that solvent and particle densities depend on $T$,
so $\Delta \rho$ is only minimized for one particular temperature.

If a colloidal sample is stirred, the initial the volume
fraction can be fairly homogeneous, and some time is needed to reach the equilibrium volume fraction gradient.  This amount of time can be estimated from the sedimentation velocity.
The Stokes drag force on a sphere moving with velocity $v$ is

\begin{equation}
F_{\rm drag} = 6 \pi \eta a v.
\label{stokesdrag}
\end{equation}

\noindent The gravitational force acting on a colloidal particle
is given by

\begin{equation}
F_{\rm grav} = \frac{4}{3} \pi a^3 \Delta \rho g .
\label{gravforce}
\end{equation}

\noindent Balancing these two gives the sedimentation velocity as

\begin{equation}
v_{\rm sed} = \displaystyle \frac{2}{9} 
\displaystyle \frac{\Delta \rho g a^2}{\eta}.
\label{sedvel}
\end{equation}

\noindent In practice, the sedimentation velocity is much slower
for high volume fraction samples due to the backflow of the solvent
through the sedimenting particles \cite{batchelor72,paulin90,he10}.
However, $v_{\rm sed}$ can be used to find a crude estimate
for relevant time scales:  the volume fraction gradient should
be established in time scales of order $z_0 / v_{\rm sed} \sim
a^{-5} \Delta \rho^{-2}$, for
example.  Measuring $v_{\rm sed}$ in a centrifuge (increasing $g$)
can be used to estimate $\Delta \rho$, again being mindful of the
temperature dependence of $\Delta \rho$.

Considerations of diffusion and sedimentation lead to
the conclusion that, all else being equal, smaller particles
are preferred.  However,  other
experimental considerations often dictate that larger particles
be used.  Where appropriate, this will be commented on in
section~\ref{techniques} which deals with experimental techniques.
Another possibility is to use colloids that are much better
density matched, and microgel particles are powerful in this
regard \cite{bartsch92,bartsch97}.  These particles are crosslinked
polymers used in a good solvent, where the particle is swollen and
permeated with solvent, and thus the density matching is much less
an issue.

\subsection{Overview of Rest of Review}

The goal of this review is to familiarize the reader with
current knowledge of properties of colloidal suspensions in
the glassy state ($\phi_g \lesssim \phi \lesssim \phi_{rcp}$)
or very near to it ($\phi \rightarrow \phi_g$).  The majority
of our attention will be given to hard-sphere-like colloids, as
many experimental and simulational results concern these systems.
We will, however, compare and contrast these observations with other
colloidal glasses, as well as with atomic and molecular glasses
when appropriate, and describe relevant theoretical attempts to
understand the nature of the glass transition.

Section~\ref{techniques} reviews experimental techniques within
the field.  It is by no means a complete review of any specific
technique, and so references will be given for further reading.
Section~\ref{approachingglasstransition} discusses what is known
about the glass transition, that is, $\phi \rightarrow \phi_g$,
and section~\ref{featuresofglasses} discusses properties of
glasses, samples with $\phi > \phi_g$.  Section~\ref{otherglasses}
discusses other soft glassy materials, and section~\ref{conclusion}
is a brief conclusion.

\section{Important Techniques}\label{techniques}

\subsection{Video Microscopy}\label{microscopy}

Microscopy has been used to study colloidal suspensions since the
work of Brown and his contemporaries, who reported on the thermal
motion of colloidal particles; a good historical account of these
observations is~\cite{haw02}.  In modern times, the
availability, commonality, and relative ease-of-use of optical
microscopes and video cameras have made video microscopy a popular
technique.  Whether used in a
biology, biochemistry, or physics setting, the mode-of-operation
is the same:  a microscope is used to visualize a system; a camera
is coupled to the microscope and is used to capture images; and,
some type of recording media stores the images for later analysis
\cite{olafsen10}.  
Probably the most familiar form of microscopy is brightfield
microscopy.  Brightfield microscopy relies on scattering or absorption of light
by the sample to produce image contrast.  Scattering occurs when
small differences in the sample's refractive index cause light to
deviate from its initial path, leading to a brightened or darkened
region in an image.  The amount of absorption depends on the
material properties of the sample, but image contrast can often be enhanced by the addition of dyes.  Modifications of brightfield microscopy include darkfield microscopy, phase contrast
microscopy, and differential interference contrast microscopy
-- all of which are effective at improving image contrast when
the variations in refractive index are small, such as the case
of a living cell (filled mostly with water) in a watery medium
\cite{olafsen10,elliot2001acis}.

Brightfield microscopy is particularly easy when the sample
is quasi-two-dimensional (quasi-2D).  For example, quasi-2D
colloidal glasses have been studied confined between parallel
plates \cite{marcus99,yunker09} or at an interface
\cite{konig05,ebert09}.  In such experiments, particles always
remain in focus and microscopy is quite easy.


A second important type of microscopy is fluorescence microscopy.
In this case, the illuminating light is high energy (short
wavelength), which excites excites a dye and causes the emission of
longer wavelength light.  The advantage of fluorescence microscopy over brightfield
is that specific constituents of a sample can be dyed,
such as particles in a colloidal suspension, and thus
selectively observed \cite{olafsen10}.  However, the
main drawback of using dyes in a sample is that they can lose
their ability to fluoresce with increased exposure to light
and oxygen -- an effect called photobleaching.  This means that, over the course of an experiment, the portion of
the sample which is being observed will become dimmer.
When studying colloidal glasses, the effect of photobleaching is
often minor; the time between successive images can be safely set
to be on the order of tens of seconds because the dynamics are slow,
minimizing the system's exposure to light.

In some cases, the presence of a dye can modify the interactions
within a system.  For example in the case of PMMA particles, some
dyes can leave a small residual electric charge on the particle,
causing them to behave as slightly soft spheres, rather than hard
ones (though, this can be countered by adding salts to the solvent
\cite{royall03,yethiraj03,yethiraj07}).  Additionally, dyes can
decay over time and, over long times, can even diffuse out of
the particles and into the solvent, making imaging difficult.

A good general discussion of video microscopy is~\cite{inoue97}.
Applications of video microscopy to colloidal suspensions are
reviewed in~\cite{murray1996arpc,olafsen10,habdas02}.


\subsection{Confocal Microscopy}\label{confocalmicroscopy}


Conventional optical microscopes are not well-suited for
three-dimensional microscopy.  In order to see deep within a sample,
it is necessary to minimize the scattering of light by closely
matching the refractive indices of the particles and solvent. Without scattering, conventional optical microscopy is difficult.
Fluorescence microscopy overcomes this by
using the contrast between dyed and undyed portions of the sample
to produce an image.  This works well for dilute samples, but is
poorly suited for dense systems such as colloidal glasses.
Because the sample
is nearly transparent, objects outside of the focal plane are
fluoresced, and stray background light passes readily through the
optics and can severely muddle an image:  it is hard to
distinguish bright particles on a bright background.  Confocal microscopes
use fluorescence as well, but overcome this limitation with
special optics (described below) and
are much better suited for studying dense colloidal systems.

\begin{figure}
\centerline{
\includegraphics*[width=0.55\textwidth]{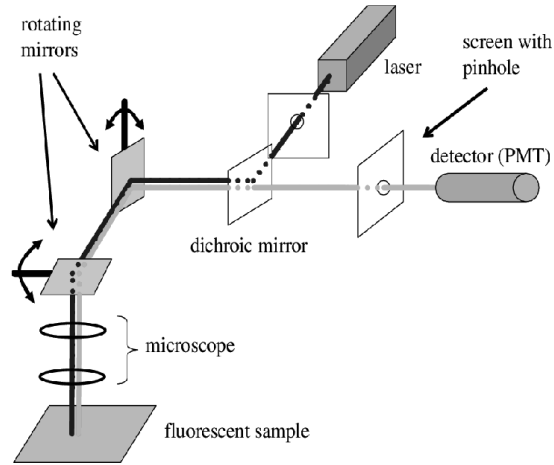}}
\caption{ (Colour online) Schematic of a confocal microscope.
Rotating mirrors scan the incoming laser light over the region
of interest in the sample.  The emitted light follows the reverse
optical path until arriving at the dichroic mirror, where it
passes through the pinhole and into the detector.  A dichroic
mirror reflects light below a certain wavelength and transmits
light above it.  Figure taken
from~\cite{prasad07} with permission.}
\label{confocalschematic}
\end{figure}

The functioning of a confocal microscope hinges on two principles:
illumination of a small sample volume ($\leq 10^{-15} L$) and
rejection of out-of-focus light~\cite{prasad07}.  A schematic of
a confocal microscope is shown in figure \ref{confocalschematic}.
Laser light, shown in black (blue online), passes through a
dichroic (dichromatic) mirror and onto rotating mirrors that
scan the light in the horizontal planes.  The light then passes
through the microscope optics and excites the fluorescent sample.
The emitted light, shown in dark gray (green online), follows the
reverse optical path back to the dichroic mirror, where it is reflected
onto a screen with a pinhole.  The pinhole is placed in the
{\it conjugate focal} plane of the sample (hence the term
{\it confocal}), rejecting the vast majority of out-of-focus
light and limiting the depth of field \cite{habdas02}.
The remaining in-focus light is finally collected by a detector,
such as a photomultiplier tube.

Confocal microscopy allows for direct imaging of a sample in two
or three dimensions.  In 2D, the pixels of an image are constructed
by scanning individual points (point scanning microscopes) or
lines of points (line scanning microscopes) over a sample.
The highest rates of scanning are achieved with use of an
acousto-optical device (AOD), in which one of the mirrors in figure
\ref{confocalschematic} is replaced with a crystal that acts as
a diffraction grating whose grating spacing can be tuned with
high frequency mechanical vibrations~\cite{draaijer88,awamura90}.
Another option is to use a Nipkow disk, which scans many points
simultaneously \cite{xiao88}; these systems can also achieve high
speeds, although more illuminated points slightly increases the
background fluorescence detected at any given point.

\begin{figure}
\centerline{
\includegraphics[width=.65\textwidth]{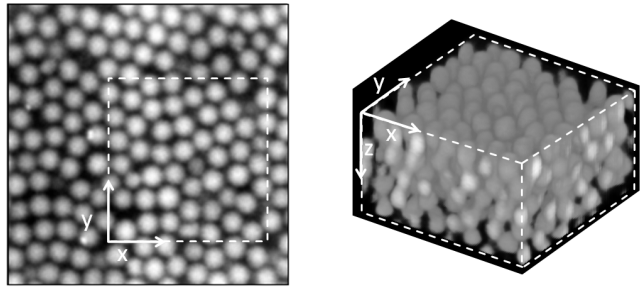}}
\caption{ Left: Confocal micrograph of a monodisperse colloidal
system at volume fraction $\phi \approx 0.63$.  The particles
have diameter $2a = 2.1$~$\mu$m and the image is taken at the
coverslip, where the particles layer against the wall.
Right: 3d reconstruction of boxed region on the left.  
Here, the image dimensions are $15 \times 15 \times
10$~$\mu$m$^3$.}
\label{3dcolloids}
\end{figure}

To obtain 3D images, such as shown in figure~\ref{3dcolloids}, the 2D scanning procedure is quickly repeated
while the focal plane is advanced through different depths in
the sample.  In the fastest modern confocals, 2D images can be
collected at rates $\approx 100$ frames/s, and depending on
the scanning depth, 3D images can be collected in around 1~s.
The specific details (times, pixels) vary from system to system,
although it is worth noting that dynamics in dense colloidal systems
are quite slow near the glass transition, so even slower confocal
microscopes can still get adequate images from glassy samples.

The earliest observations of colloids using confocal microscopy were
done by Yoshida, Ito, and Ise in 1991 \cite{ise91} and van Blaaderen
{\it et al.} in 1992 \cite{vanblaaderen92}.  Yoshida {\it et al.}
examined colloidal crystallization near walls, and later studied
colloidal gels \cite{ise94}.  van Blaaderen {\it et al.}
demonstrated the utility of fluorescent core-shell particles and
confocal microscopy.  Core-shell particles are ones with
small fluorescent cores and non-fluorescent shells, so that their
centres are bright dots that are well separated in the image from other
particle centres even at high volume fractions.  The early work
of van Blaaderen {\it et al.} nicely demonstrated the power of
confocal microscopy with important proof-of-principle
measurements, and hinted at applications using particle tracking
\cite{vanblaaderen92,vanblaaderen93}.  In 1995 van Blaaderen and
Wiltzius applied particle identification software to locate the
positions of several thousand particles in 3D confocal images to
investigate the structure of a colloidal glass
\cite{vanblaaderen95}, sparking much subsequent work
\cite{kegel00,weeks00}.  The key 1995 finding was that the
structure of a colloidal glass was quite similar to glassy
structure seen in simulations.

More details of applying confocal microscopy to colloidal
samples can be found in~\cite{prasad07,chestnut97}, and
a good starting point to learn about confocal microscopy is~\cite{pawley06}.

\subsection{Particle Tracking}\label{tracking}

Particle tracking incorporates various image processing and computational techniques to identify the centroid positions of particles in a given image~\cite{habdas02,crocker96}.  Images can be two-dimensional, as in brightfield or fluorescence microscopy, or three-dimensional, as in confocal microscopy.  Repeating the procedures for consecutive images yields a list of coordinates at subsequent times.  The coordinates can be used immediately to obtain structural information about a sample, or if dynamic information is desired, the coordinates can be linked together in time to form individual particle trajectories.

In general, the larger a particle is in an image, and the more it contrasts with the background, the more accurate the particle tracking.  As mentioned in section~\ref{basicphysics}, however, larger particles move slower and are more prone to sedimentation.  For many experiments, particle centres can be located with a resolution of approximately 20 nm in the focal plane, while the out-of-plane resolution is typically no better than 50 nm.  Recently, algorithms have been developed that push spatial resolution to $\approx$~5 nm~\cite{kilfoil2009,jenkins2008acis}.

In dilute samples, accurately identifying particles is relatively easy because bright and well-separated particles contrast well with the dark background.  In dense samples like colloidal glasses, there are many bright particles in an image and so contrast is usually poorer.  Additionally, optical effects such as diffraction can make it difficult to distinguish individual particles when they are very close together.  These effects are important to understand and correct, especially when particle motions are very small~ \cite{crocker99,baumgartl05}.
To illustrate, in a sample of 2.4 $\mu$m PMMA spheres at $\phi = 0.52$, Weeks and Weitz observed the majority of particles to move less than 0.2 $\mu$m over 600 s~\cite{weeks02}.  The influence of diffraction can be weakened by increasing the optical resolution by using fancier lenses~\cite{inoue97}, by using confocal microscopy (see section~\ref{confocalmicroscopy}), or with computational techniques~\cite{jenkins2008acis,crocker99}.  Hence, with some care as far as optics are concerned, and some fine tuning of particle tracking parameters, it is often straightforward to study dilute and dense systems with the same techniques.

Combined with video microscopy, particle tracking offers a powerful method to probe the local properties of a sample, which is especially important for understanding structurally or dynamically heterogeneous systems like colloidal glasses. With this technique, one can discuss behaviours of individual particles up to a collection of several thousand.  This degree of resolution is not available with light scattering (see section~\ref{dls}) or conventional rheology (see section~\ref{rheology}) where quantities are averaged over thousands to millions of particles.  However, such a small statistical sampling can make it difficult to draw conclusions about a system's bulk properties without collecting an overwhelming amount of data.

The main starting point for particle tracking is the original article by Crocker and
Grier~\cite{crocker96}, and the software described in the article is available for download on the web~\cite{idlwebsite}. Samples that are flowing or being sheared can also be tracked using pre-treatment
of the data; see~\cite{besseling09} for details.  For a comprehensive assessment of particle tracking, see~\cite{jenkins2008acis}. 

\subsection{Static and Dynamic Light Scattering}\label{dls}

Light scattering is a powerful technique for probing the average
structure and dynamics of a sample.  A laser is aimed at a sample,
and the light scattered from the sample at a given angle is detected.

Photons scattered from different portions of the sample interfere
with each other, and how this interference (constructive or
destructive) depends on angle provides information about the
structure of the sample.  In particular, this information leads
to the static structure factor $S(k)$, the Fourier transform of
the particle positions.  This is static light scattering (SLS).
The scattering wave vector $k$ is given by $k = \left[ 4 \pi n /
\lambda \right] \sin(\theta/2)$, where $\lambda$ is the laser
wavelength, $n$ is index of refraction of the sample medium,
and $\theta$ is the angle between the incident light and detected
light \cite{prasad07,berne76,jones91}.

\begin{figure}
\centerline{
\includegraphics[width=.65\textwidth]{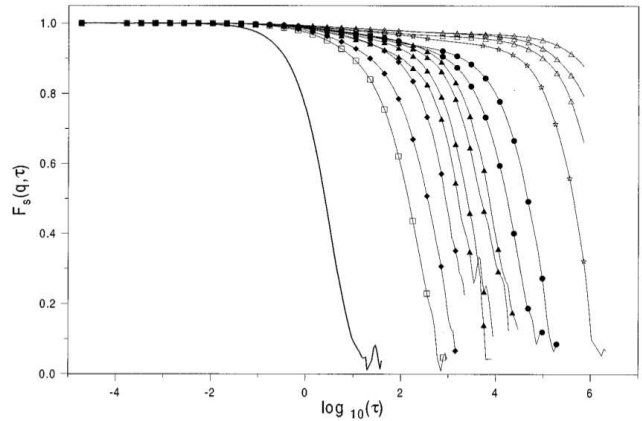}}
\caption{Self-intermediate scattering functions, $F_s(k,\tau)$,
as a function of the dimensionless time $\tau = \Delta t /
\tau_D$ with $\tau_D = 0.0215$~s.  Data and symbols are the same as
figure~\ref{examplemsd}(b).  Note that an increase in $\phi$ corresponds to an increase in decay time.
Figure reprinted from~\cite{vanmegen98}. {\bf permission has been requested}.
}
\label{dlsfig}
\end{figure}

In dynamic light scattering (DLS), the light intensity, $I(t)$,
at a fixed angle is monitored as a function of time.  The light
intensity fluctuates as portions of the sample rearrange, changing
the interference pattern of scattered light.  In particular, one monitors how the intensity autocorrelation function,
\begin{equation}\label{g2}
g_2(\Delta t) = \frac{\langle  I(t+\Delta t)I(t) \rangle _t}{\langle I(t) \rangle ^2},
\end{equation}
changes as a function of lag time $\Delta t$.  At $\Delta t = 0$, $g_2(\Delta t)$ is at a maximum, and decays from this value as the
sample evolves. Scattering functions, such as the self-intermediate scattering function shown in figure~\ref{dlsfig}, are related to $g_2(\Delta t)$ and are used to quantify dynamics.
By measuring the rate of decay, one measures how particles
move and can extract information similar to the diffusion coefficient.
Probing the dynamics at different $k$ values allows one to determine
information about either local or collective particle motion within
the sample; most typically, $k$ is chosen to coincide with the
peak of the structure factor $S(k)$, which yields information about
collective motions.  Alternatively, tracer systems can be prepared
and single-particle motion probed \cite{vanmegen86,vanmegen98},
or else the behaviour at $k \rightarrow \infty$ can be examined
which also relates to self-diffusion \cite{snook83}.  For example,
the MSDs in figure~\ref{examplemsd}(b) are calculated from the DLS
data in figure~\ref{dlsfig}.  The autocorrelation function is often
calculated over time scales down to $10^{-6}$~s, allowing a large
range of time scales to be measured, as shown in figure~\ref{dlsfig}.
For more details about both SLS and DLS, see~\cite{berne76,jones91}.

The main strength of light scattering is that light is
scattered from a significant volume within the sample, typically
containing millions of particles.  The measurement is a very
good average of information from all of the particles, whether
it be structural information (SLS) or dynamic information (DLS).
For DLS, given that the measurement is sensitive to motions
corresponding to fractions of $\lambda$, accurate
MSDs are straightforward to obtain.  However, because of the
ensemble-averaging properties, local information is harder
to obtain.  For example, while calculating the MSD is easy,
knowing how individual particle motions are correlated in space is
more difficult.  A secondary strength of light scattering is that
typically particles smaller than those in microscopy experiments
can be used, 300 nm radius~\cite{snook91} in one
early experiment.  These smaller particles are much less affected
by sedimentation (see section~\ref{basicphysics}).  Additionally,
smaller particles diffuse faster, allowing their motion to be
probed over a larger range of time scales.

In 2001, Williams and van Megen devised a clever method to examine
binary samples (mixtures of two particle sizes) with SLS and DLS
\cite{williams01}.  They slightly modified the synthesis method
for the two particle types so that they had distinct indices
of refraction.  By tuning the temperature of the solvent, they
could closely match the index of one or the other particle type,
and get information about each particle species independently.
These experiments nicely demonstrated that at a given volume
fraction, mixtures of two sizes are more liquid-like than a
monodisperse sample \cite{williams01}, in agreement with viscosity
measurements \cite{hoffman92,mewis94}.  This is because binary
samples can be packed to higher volume fractions than monodisperse
samples, so at a given volume fraction, a binary sample has more
free volume than a monodisperse sample.  A subsequent experiment
suggested that the small particles can ``lubricate'' the motion
of the large particles \cite{lynch08}.  Binary samples in general
are of interest for understanding multi-component molecular
glasses, and the technique of Williams and van Megen demonstrates
how light scattering can be used to study such multi-component
systems \cite{williams01}.

In very dense colloidal suspensions, such as those near a glass
transition, additional experimental issues arise.  One problem
is that in a glassy sample, particles don't rearrange
significantly, so it is difficult to get a proper average from
the sample.  Several techniques have been developed to deal with
this situation and are reviewed in~\cite{pusey89,cipelletti02,scheffold07}.

A second problem is that light is often scattered from more than one particle before being
detected.  Again, several techniques have been developed for
cases when the light scattered a few times before detection,
some of which are reviewed in~\cite{scheffold07,pusey99}.
One common technique is diffusive wave spectroscopy (DWS)
\cite{maret87,wolf88,akkermans88,pine1988prl,pine1990jpfrance}
which works when light scatters many times before detection.
Here, the light is assumed to be scattered so many times that
each photon can be thought to diffuse randomly through the sample
before exiting and being detected.  Diffusion is straightforward to
describe mathematically so, for a given experimental geometry, it
is possible to calculate the average number of times a photon has
been scattered (and therefore the number of particles from which
it scattered).   Again, the intensity of light is monitored and
its autocorrelation calculated, but now, each particle needs only
move a fraction of a wavelength before the sum of these motions
results in significant decorrelation of the intensity. DWS is
thus useful for multiply scattering samples with small motions.  Colloidal glasses were studied soon after the development of DWS~\cite{pine1988prl}, an early result being that the MSD of densely packed particles is nonlinear in time
(as shown in figure~\ref{examplemsd}, for example).

DWS is reviewed in~\cite{maret97}.  A useful review article which
briefly discusses differences between DWS and DLS is~\cite{klein96}.
Ultra-small-angle neutron and x-ray scattering as applied to
colloidal glasses is reviewed in~\cite{bhatia05}; these techniques
can probe structure on length scales of $\sim 1-10$~$\mu$m.
A recently published book on glasses and dynamical heterogeneity
(see section~\ref{dynhet}) contains a chapter by Cipelletti and
Weeks which focuses on colloidal glasses \cite{cipelletti11book}.
This chapter discusses many details of light scattering.  A
review article by Sciortino and Tartaglia compares experimental
data with theoretical predictions, with a focus on light
scattering data \cite{sciortino05}.

\subsection{Rheology}\label{rheology}

Rheology is the study of how materials flow and deform.
A rheological measurement quantifies how solid- or fluid-like a
substance is in response to a specific stress \cite{larson98};
that is, the goal of rheology is to measure elastic and viscous
properties of a system.  To make such a measurement, one needs
a \textit{rheometer}, a device capable of either creating a
constant or oscillatory stress and measuring the resulting rate of
deformation, or measuring the stress required to deform a material
at a constant rate of strain.

\begin{figure}
\centerline{
\includegraphics[scale=.6]{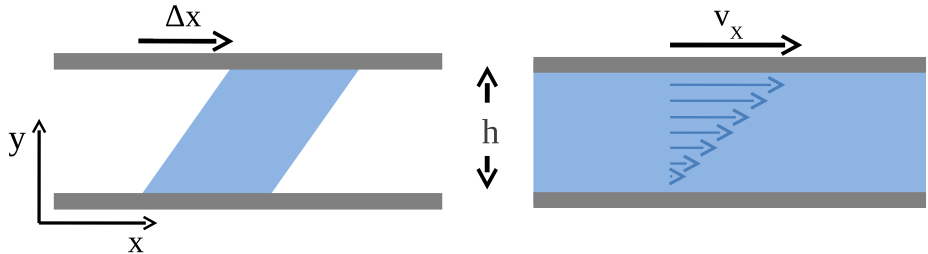}}
\caption{ (Colour online) Schematic of a simple rheometer.  Left,
the top of an ideal solid is displaced $\Delta x$, leading to
a strain of $\gamma = \Delta x/h$.  Right, a Newtonian fluid is
sheared at a rate of $\dot{\gamma} = v_x/h$.}
\label{shearschematic}
\end{figure}

An elementary rheometer is illustrated in figure \ref{shearschematic}.
The device consists of two horizontal plates separated by a
distance $h$, where the top plate is mobile and the bottom plate
is fixed.  On the left of figure~\ref{shearschematic}, an ideal Hookean
(elastic) solid is placed inside and the top plate is displaced by
a distance $\Delta x$.  The stress $\sigma$ ($\equiv$ force/area)
needed to do this is given by the relation

\begin{equation}
\label{hookeansolid}
\displaystyle \sigma = G \frac{\Delta x}{h} = G \gamma.
\end{equation}

\noindent This equation defines the {\it shear
modulus} $G$, where $\gamma = \Delta x/ h$ is the strain.  In this
case, the stress depends only on the fixed material quantity $G$
and the magnitude of $\Delta x$.

On the right side of figure~\ref{shearschematic}, the rheometer is filled with a
simple fluid, such as water, and the top plate is
displaced at a constant velocity $v_x$.  As the top plate moves,
it drags the fluid underneath in accordance with the no-slip
boundary condition of fluid mechanics.  For the same reason,
the fluid immediately above the fixed bottom plate is motionless.
This creates a steadily decreasing velocity profile (indicated by
the arrows).  The shear stress needed to maintain the constant
velocity of the top plate is given by

\begin{equation}
\label{newtonshear}
\displaystyle \sigma = \eta \frac{\partial v_x}{\partial y} = \eta \frac{v_x}{h} = \eta \dot{\gamma}.
\end{equation}

\noindent The above equation defines the {\it shear
viscosity} of a fluid, $\eta = \sigma /  \dot{\gamma}, $ where
$\dot{\gamma}$ is the shear rate.  Fluids that adhere to this
relation are called {\it Newtonian fluids}.

\begin{figure}
\centerline{
\includegraphics[width=.5\textwidth]{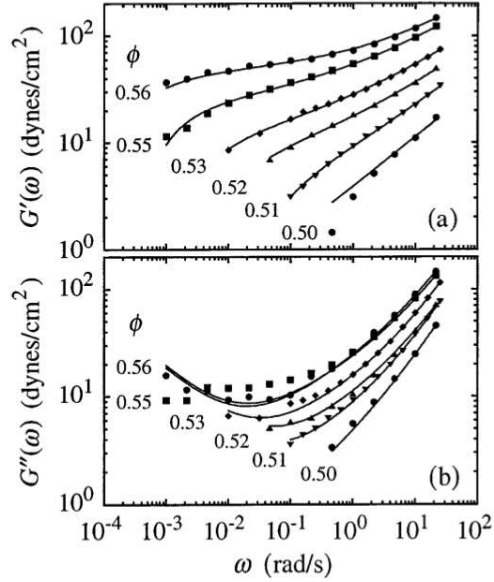}}
\caption{ 
(a) Storage modulus and (b) loss modulus as a function of
frequency for colloidal suspensions at different volume fractions.  The solid lines are fits to a model based on mode-coupling theory.
Figure reprinted from~\cite{mason95glass} -- {\bf permission has been requested.}
}
\label{masonfig}
\end{figure}

More generally, many materials are termed {\it
viscoelastic}:  they have both a viscous and elastic nature
\cite{zaccarelli2009jcp,mohan2010sm,puertas2005jpcm,pham06,mason95glass,conrad06}.
Viscoelasticity can be studied by applying a low amplitude sinusoidally varying strain of the form $\gamma = \gamma_0 \sin(\omega t)$. As noted above, the elastic stress is proportional to this strain and thus depends on $\sin(\omega t)$, while the viscous stress is proportional to the strain rate and thus depends on $\cos(\omega t)$.  For a viscoelastic material one would measure

\begin{equation}
\sigma (t) = \gamma_0 \left[G'(\omega) \sin (\omega t) + G''(\omega) \cos (\omega t) \right],
\end{equation}

\noindent where the two moduli are the {\it storage modulus}
$G'(\omega)$, and the {\it loss modulus} $G''(\omega)$.  These
two moduli in general depend on the frequency $\omega$.
$G'$ describes the ability of the material to store elastic energy,
while $G''$ characterizes energy dissipation.  Analogous to a
dampened spring, the elastic portion oscillates in phase with the
stress, whereas the viscous portion is out of phase by a factor
of $\pi/2$.

It is worth noting that colloidal suspensions are viscoelastic,
and so their rheological properties depend on the measurement
frequency, $\omega$, as shown in figure~\ref{masonfig}~\cite{mason95glass}.  Viscoelastic behaviour
has been explored both experimentally and theoretically
\cite{mason95glass,schepper1993prl}. In figure~\ref{masonfig}, it can be seen that both $G'$ and $G''$ rise rapidly near the glass
transition over a large range of of $\omega$.  (See the
discussion of sheared samples in section \ref{shear}).

For colloids at the glass transition, the
elastic modulus $G'(\omega)$ is larger than the loss modulus
$G''(\omega)$ for a wide range of frequencies, and in particular
as $\omega \rightarrow 0$.  This latter condition corresponds to
solid-like behaviour for a quiescent sample.  Related to this is the
idea of a yield stress, that a solid-like sample requires a finite
stress be applied in order for the sample to flow (flow being defined
as $\dot{\gamma} > 0$ for a given applied stress)
\cite{pham06}.

There exist techniques to measure viscosity and elasticity from
video microscopy and particle tracking, and light scattering;
these techniques are collectively termed {\it microrheology}
\cite{mason95micrheo,mason97,crocker00,breedveld03,waigh05}.  It is
important to note that microrheology measurements represent local,
microscopic properties, whereas rheological measurements involve
macroscopic, bulk samples.  Applying microrheology methods to dense
suspensions requires care in the interpretation~\cite{habdas04,
squires05, carpen05, drocco05, williams06, meyer06, reichhardt06, kaufman07,
gazuz09, khair10,wilson11}.  Of historical note, the first
experiment to use microrheology was by Mason and Weitz in 1995, and
this experiment used DWS (section~\ref{dls})
to probe colloidal glasses as a test case.

Good reviews about the rheology of colloidal suspensions
include~\cite{bergenholtz01,lionberger01,stickel05,buscall10}.
More general information about rheology can be found
in~\cite{larson98,macosko94,jones02,chen10rheology}.

\subsection{Simulations}\label{simulations}

As discussed above, colloidal glasses are often considered as model hard-sphere glasses, and complement simulations
of hard-spheres.  Likewise, simulations of hard-spheres give quite
useful insight to colloidal glasses, and in many cases have guided
experiments.

It is difficult to simulate a large number of colloidal particles at
high volume fractions taking into account hydrodynamic interactions
and interaction potentials; often approximations are desirable or
necessary \cite{brady88}.  For that matter, colloidal glasses are
themselves only approximate models of molecular glassy materials,
so to the extent that colloidal glasses may provide
insight into the general glass transition, one hopes that the
details are not crucial and that approximations are acceptable.
Fortunately, this seems to be the case.  First, the microscopic
dynamics seem unimportant.  Simulations with
Brownian dynamics (appropriate for colloids) or Newtonian dynamics
(appropriate for simple hard-sphere systems without a solvent) result in similar long-time-scale
dynamics \cite{binder98,szamel04,tokuyama03,tokuyama07,berthier07a,franosch08}.
Second, the interaction potential seems unimportant.  Observations
such as dynamical heterogeneities are similar in Lennard-Jones
simulations \cite{kob97,poole98,donati99,berthier04}, hard-sphere
simulations \cite{doliwa98,doliwa00}, and soft sphere simulations
\cite{hurley95,hurley96,yamamoto98,kurita10b}.  The accumulation
of evidence suggests that the specific details of colloidal
interactions are not crucial for understanding glassy behaviour.
The limitations of the colloidal samples as models (Brownian
dynamics with hydrodynamic interactions) may also not be crucial
problems for comparing colloidal glasses with molecular glasses.
Third, even the dimensionality may be fairly unimportant.
Simulations see similar slowing of dynamics in 2D and 3D, as well as similar particle motions~\cite{doliwa00,widmercooper04,donati98}.  Likewise, colloidal experiments
see similar slowing and similar qualitative features in 2D
\cite{marcus99,konig05} and 3D \cite{kegel00,weeks00,bartsch98}.
One caveat is that preventing ordering is more important in
lower dimensions and so binary or polydisperse samples must be
used to study glass transitions in 2D.  However, this also
suggests the possibility of better understanding the role of
crystallization and frustration by considering higher dimensions;
see~\cite{vanmeel09a,vanmeel09b,charbonneau10} which discuss
intriguing results from four-dimensional simulations.

Another consideration  for comparing simulations and
experiments are finite size effects.  Simulations are most directly
comparable to microscopy experiments.  In a simulation, often
periodic boundary conditions are used.  The key assumption, then,
is that the box size should be at least twice as big as 
any structural length scales or dynamical length scales present
\cite{andersen05}.  Of course, it is possible some of these
length scales may be longer than expected -- for example, one
simulation found evidence for a structural length scale that was
three times as large as the more obvious dynamical length scale
\cite{kob00}.  Two simple options exist.  First, one can conduct
simulations for a range of box sizes, and verify that the physics
one observes is independent of box size or perhaps scales
in some clear way.  Second, one can exploit the
size dependence to learn something about the sample
\cite{yamamoto00}; see the discussion in section~\ref{confine}.

In experiments, finite size effects also can cause problems.
In a typical microscopy experiment, 2D images can contain a few
hundred particles, or 3D confocal microscopy images can contain
a few thousand particles.  While the sample chamber may well be
much bigger, this still limits the size of dynamical length scales
that can be studied; see for example the discussion of finite size
effects in~\cite{weeks00}.  Also, samples are often imaged
through a glass coverslip, and care must be taken to take the data
away from the boundaries; the presence of boundaries introduces
layering \cite{lowen99,murray98,archer07,nugent07prl,desmond09} and
likely changes the dynamics as well \cite{lowen99,edmond10b}.
However, microscopy imaging is difficult deep within a sample;
here light scattering has an advantage.

Given the similarities between a variety of simulations and the
colloidal glass transition, this review article will not completely
survey the literature of simulational studies of the colloidal
glass transition because in reality, simulations can be quite
relevant for the colloidal glass transition without specifically
being simulations of colloids, and this review article cannot
effectively survey all of the simulations of the glass transition.
Instead, in subsequent sections of this review, as we describe
features of the colloidal glass transition we will discuss the
relevant simulation results, and we will use the cases to compare
the strengths and weaknesses of experiments and simulations.

However, certainly some intriguing advantages of simulations are worth
noting here.  Widmer-Cooper {\it et al.} demonstrated advantages
of the ``iso-configurational ensemble,'' where they repeated
simulation runs with identical starting positions for particles,
but with randomized velocities; this is certainly something
well-suited to simulation \cite{widmercooper04,widmercooper05}.
Their results are described more fully in section~\ref{dynhet},
but briefly, their technique demonstrated that certain regions
have a higher propensity for particle motion.  Another
interesting simulation by Santen and Krauth used non-physical
Monte Carlo moves to probe ``equilibrium''-like sample properties
for glassy samples \cite{krauth00}.  They found that
thermodynamic properties were continuous across the transition,
evidence that the glass transition is not a thermodynamic
transition.  The four-dimensional simulations mentioned above led to interesting results as crystallization is much harder
in 4D, and so even a monodisperse system can have glassy behaviour
\cite{vanmeel09a,vanmeel09b,charbonneau10}.  These three examples
-- all using hard particles -- give a sense of the variety of
ways simulations can give unique insight into the colloidal glass
transition.

Several textbooks exist which discuss simulation techniques; a
good starting point is~\cite{frenkel01}.  Reviews of simulations
of the glass transition include~\cite{glotzer00,andersen05}.

A textbook introducing a large variety of methods for studying
soft materials is~\cite{olafsen10}.  Many of the
techniques discussed above are described in more detail,
including microscopy, simulation methods, and rheology.

\section{Features of Systems Approaching the Glass Transition}
\label{approachingglasstransition}

\subsection{Growth of Viscosity and Relaxation Times}
\label{divergence}

\begin{figure}
\centerline{
\includegraphics[width=.7\textwidth]{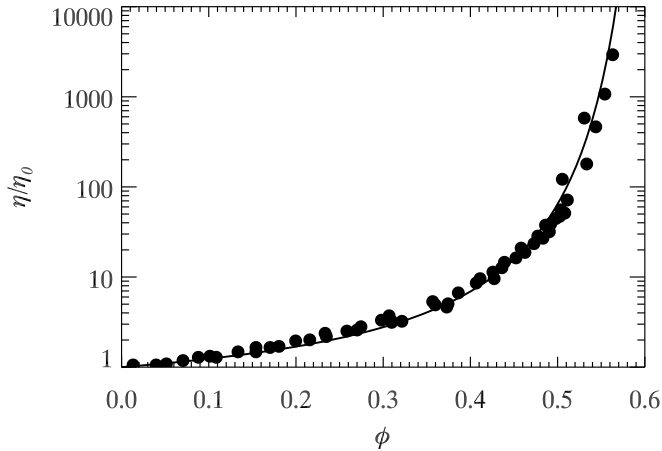}}
\caption{ Scaled low shear viscosities at different $\phi$ for
various colloidal suspensions of nearly monodisperse hard-spheres.
The low shear viscosities ($\eta$) are normalized by the viscosity
of the pure solvent ($\eta_0$).  
The fit line is to Eqn.~\ref{eqn:doolittle} with $C = 1$, $\phi_m = 0.638,
D = 1.15$.
Data taken from~\cite{cheng02,phan96pre,segrepusey97,russelrheo}.  
 }
\label{chaikinviscosity}
\end{figure}

\begin{figure}
\centerline{
\includegraphics[width=.5\textwidth]{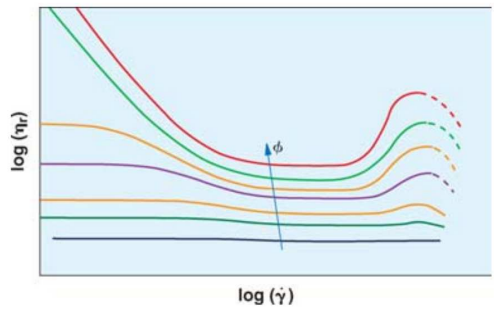}}
\caption{ Schematic of shear-thinning and shear-thickening for
colloidal suspensions at various volume fractions.  Figure reprinted from~\cite{stickel05} -- {\bf permission has been requested.}
}
\label{stickelfig}
\end{figure}

A liquid's viscosity increases upon cooling.  If cooling continues
into the supercooled regime, the viscosity continues to grow, and
at the glass transition is about $10^{13}$ poise \cite{angell95}.
(For comparison, the viscosity of water at room temperature is
0.01~P, glycerol is 15~P, and honey is 100~P \cite{habdas06}.)
Analogously, increasing the volume fraction in a colloidal
suspension, shown in figure~\ref{chaikinviscosity}, causes an
increase in viscosity.  As can be seen, the maximal change in
viscosity is only a factor of $10^4$; indeed, one critique of
the colloidal glass transition as a model for molecular glass
transitions is that the viscosity increase is not nearly as great.
This discrepancy likely arises for several reasons.  First, it
is experimentally difficult to load high volume fraction samples
into a rheometer~\cite{poon12,cheng02}.  Such a limitation can
potentially be overcome by using thermosensitive particles, which
could be loaded into the rheometer at a temperature where the
sample is liquid-like and then thermally changed to a higher volume
fraction {\it in situ} \cite{crassous08,siebenburger09}.  Second,
ensuring that the sample has a well-known and controllable volume
fraction can be extremely challenging~\cite{poon12}.  Finally,
colloidal samples that are sheared too rapidly can shear thin
(the apparent viscosity decreases with increasing shear rate) or,
at still higher shear rates, shear thicken (an increasing apparent
viscosity with increasing shear rate).  These trends are indicated
qualitatively in figure~\ref{stickelfig}.  Shear thinning is more
severe for $\phi > 0.5$ \cite{marshall1990,yamamoto97}, meaning that
experiments at high $\phi$ must be done at extremely low shear rates
($\omega \rightarrow 0$ as described in section~\ref{rheology})
and low applied stresses to see the correct linear response
\cite{williams06}.  Measurements  for $\phi \approx 0.6$ would
take weeks or years to be done properly~\cite{cheng02}.

Important early work on the viscosity of colloidal
suspensions was performed by Marshall and Zukoski using
rheometry \cite{marshall1990}.  Their system consisted of
small silica hard-spheres (radius $< 300$~nm) in a solvent
of decahydronapthalene. A constant stress rheometer was used
to measure viscosity at various applied stresses, enabling an
extrapolation of the viscosity to a state of zero stress.  For all
particle sizes used, they observed an increase in viscosity with $\phi$, with a sudden, apparently divergent increase at
volume fractions associated with the glassy phase of hard-spheres.
They also found that the form of the increase was well-described
by the Doolittle equation,

\begin{equation}
\displaystyle \frac{\eta}{\eta_0} = C \exp{\left[\frac{D \phi}{\phi_m - \phi}\right]},
\label{eqn:doolittle}
\end{equation}

\noindent 
with $C=1.20, D=1.65$, and $\phi_m=0.638$.  This equation
was first used to describe the temperature dependence
of viscosity in molecular liquids approaching the glass
transition~\cite{doolittle1951}.  The original Doolittle equation
was expressed as a function of free-volume (which was implicitly
a function of temperature).   We note that (\ref{eqn:doolittle})
has been modified in a reasonable fashion (see~\cite{marshall1990}
for details) to depend on $\phi$ as shown above, with $\phi_m$
being the maximum packing.

At $\phi = \phi_m$ in the above equation, the viscosity
diverges.  Interestingly, in the glassy regime, ($\phi \geq
0.52$ for these data), the value $\phi_m = 0.638$ gave a
remarkably good fit to the majority of the data and is very
close to $\phi_{rcp}$, where all motion is suppressed.
Indeed, the data in figure~\ref{chaikinviscosity} are also
well-fit by (\ref{eqn:doolittle}) with a similar $\phi_m$
\cite{cheng02}.  It is surprising that the divergence
is at $\phi_{rcp}$ rather than $\phi_g \approx 0.58$~\cite{cheng02}.
This raises questions about exactly what occurs at $\phi_g$ and
the utility of colloidal glasses as models for molecular glasses.
The experiments of \cite{cheng02} are only able to measure a
change in viscosity of four orders of magnitude, far less than is seen for molecular glasses.  It is conceivable that the
experimental $\phi_g \approx 0.58$ is rather far from a more true
glass transition point for colloids -- perhaps at a volume fraction
comparable to $\phi_{rcp}$.  To the extent that one cares
about properties of molecular glasses extremely close to $T_g$,
it would be disappointing if dense colloidal liquids can only
be equilibrated in samples relatively far from their true $\phi_g$.  We take the view that, while one
should be aware of these possible limitations, colloids are still a useful model system for understanding the
glass transition.  As noted in section~\ref{simulations},
the agreement between colloidal experiments and computational
simulations strengthens the validity both.  (Simulations are also somewhat
limited in the time scales that they can address~\cite{glotzer00}.)
Furthermore, over the volume fraction range for which samples
can be equilibrated ($\phi \lesssim 0.6$), colloids share
many similarities with molecular glasses, both in experiments and
simulations.  These similarities strengthen the utility of the colloidal
model system, despite the relatively limited viscosity range that
is observable.

The Doolittle model has been critiqued in the past
as being oversimplified or perhaps founded on shaky physical
arguments \cite{tarjusviscous,angell1981}, and it is possible
that other functional forms would fit the data just as well
\cite{puertas2005jpcm,cheng02}.  The question of which functional form is most appropriate is generic to studying the glass
transition.  It was noted by Hecksher {\it et al.} in 2008 that
multiple functional forms fit glass transition data (relaxation
times as a function of $T$). Of these expressions, some
have a divergence at finite $T$ while others have no divergence at all~\cite{hecksher08,mckenna2008nat}.  In all cases, the
experimental data range over many decades in $\eta$, but are clearly
many more decades away from $\eta = \infty$, and so extrapolation
is always tricky \cite{teitel2007,olsson11,vaagberg11b}.


While the glass transition is associated with a dramatically
increased viscosity, it is equally associated with a dramatically
increased microscopic relaxation time and decreased diffusivity.
For colloids, the long-time self-diffusion coefficient
$D_L(\phi)$ approaches zero as $\phi \rightarrow \phi_g$ (see
section~\ref{basicphysics} for discussion of $D_L$).  A related
quantity is the intermediate scattering function $F(k_m,\tau)$,
where the wave vector $k_m$ is often chosen to correspond to the
peak of the static structure factor.  The decay
time for $F(k_m,\tau)$ becomes large as the glass transition
is approached, as shown in figure~\ref{dlsfig}; this is the microscopic relaxation time scale,
often termed $\tau_\alpha$ when referring to the final decay of
$F(k,\tau)$~\cite{ediger00}.  (See section~\ref{dls} for discussion of dynamic light scattering (DLS) and scattering
functions.)  Roughly, $\tau_\alpha \propto a^2/D_L$, where $a$
is the particle radius, and so both $\tau_\alpha$ and $D_L$ are
considered measures of how microscopic dynamics slow near the
glass transition.

\begin{figure}
\centerline{
\includegraphics[width=.7\textwidth]{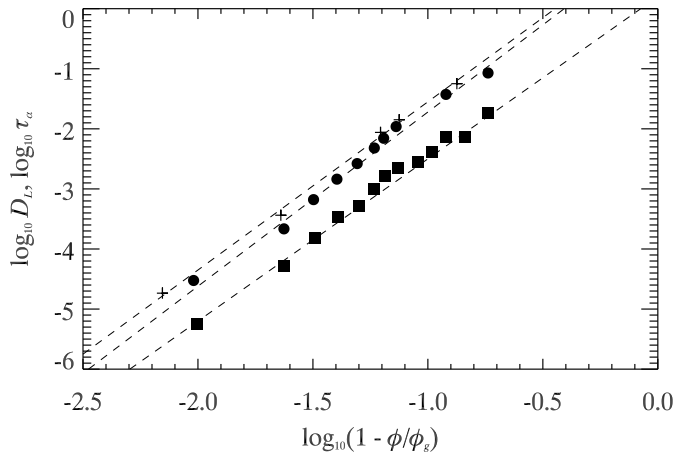}}
\caption{
Growing relaxation time scale $\tau_\alpha$ (plotted as
$-\log (\tau_\alpha)$, circles)
and decreasing diffusivity $D_L$ (squares) plotted against the distance to
the colloidal glass transition, using $\phi_g = 0.571$ for the
$D_L$ data and $\phi_g=0.572$ for the $\tau_\alpha$ data. Data taken from~\cite{vanmegen98}.
Diffusion constants are measured from figure~\ref{examplemsd}(b)
and $\tau_\alpha$ from figure~\ref{dlsfig}.
Plus symbols are $\tau_\alpha$ data from \cite{vanmegen94}, a
prior experiment by the same group.
}
\label{growingtau}
\end{figure}

The question then is how $D_L$ and $\tau_\alpha$ depend on $\phi$
\cite{ngai98,ngai07}.  Results from viscometry and DLS studies
were reported by Segr\`{e} \emph{et al.} for a suspension of PMMA
hard-spheres over the range of volume fractions $0 \leq \phi \leq
\phi_{\rm freeze} = 0.494$ \cite{segre1995prl}.  Intriguingly, they
found that the growth of $\eta(\phi)$ (measured with a rheometer)
was well-matched by the growth of the inverse diffusion constant,
$[D_L(k_m)]^{-1}$.
This suggests that at least to $\phi \approx 0.494$, viscosity
and diffusion are well-coupled; it is known that in molecular
glasses, these two quantities can decouple with diffusion occurring
slowly, but not as slowly as would be expected from measurements of $\eta$~\cite{ediger00,fujara92,ediger93,chang94}.  
van Megen {\it et al.} acquired data up to $\phi_g - 0.01$ and
found that $D_L$ and $\tau_\alpha$ remain well-coupled; their data
are shown in figure~\ref{growingtau}.
Results for both
$\eta$ and $\tau_\alpha$ from the same sample with $\phi > 0.494$
have not yet been obtained, partly because of the experimental
difficulty of making the different measurements at exactly the
same volume fraction \cite{poon12,segre95comment,segre95reply}.
Comparing data sets from different groups suggests that perhaps
$\eta$ and $\tau_\alpha$ remain coupled \cite{bonn03}, although
such comparisons are non-trivial and the results should be
treated cautiously \cite{poon12}.
In general, it is hard to accurately determine how $\eta(\phi)$
grows near $\phi_g$ or $\phi_{rcp}$, partly because of the
difficulty in measuring $\phi$ accurately, and partly because small
changes in $\phi$ make a large difference, precisely as shown
in figure \ref{chaikinviscosity} and figure \ref{growingtau}.
It is worth noting that some experimental differences between
$\eta(\phi)$ and $D_L(\phi)$ may be due to slip at the particle
surface~\cite{segre1995prl,imhof94}.  As already noted, another
difficulty in measuring $\eta(\phi)$ for $\phi>0.5$ arises due to
the very slow shear rates required to do so -- the need to shear
at such low rates is itself evidence for a dramatically growing
microscopic relaxation time scale \cite{marshall1990}.

Given the difficulties of doing both viscometry and DLS on the
same samples, and given the power of DLS compared to viscometry,
it is natural that many people have used DLS to examine how
$\tau_\alpha$ grows as $\phi \rightarrow \phi_g$.  The van Megen
group has performed well-known DLS experiments over three decades
starting in the 80's \cite{vanmegen86,pusey1987,vanmegen88,vanmegen89,
vanmegen91,vanmegen94,vanmegen98,vanmegen09}.  One of their notable
findings is that the increase of $\tau_\alpha(\phi)$ is well-described by mode coupling theory
(MCT); see for example~\cite{vanmegen94,gotze91,sciortino05}.
Significantly, not only is $\tau_\alpha$ fit by MCT, but several
other features of $F(k,\tau)$ are as well, with the only adjustable parameter being the scaling of the volume fraction $\phi$.
These experiments were performed with hard-sphere-like colloidal
particles; parallel experiments were performed by Bartsch,
Sillescu, {\it et al.} starting in the 1990's, using softer
colloidal particles \cite{bartsch92,bartsch93,bartsch95,bartsch97,
bartsch02,eckert02,eckert03,eckert04,willenbacher11}.  Some
intriguing differences were seen that were attributed to the
particle softness~\cite{bartsch97}. For example, some relaxation
processes still appeared to persist in the glassy phase, perhaps
due to depletion effects (see discussion in \cite{bartsch02} and
section~\ref{softparticles}).  Furthermore, the glass transition
appeared at $\phi \approx 0.64$, close to random close packing.
However, many of the predictions of MCT were still confirmed in
these experiments \cite{bartsch97}, a solid result suggesting that
the particle properties are not crucial.

The hard-sphere results have been recently updated by Brambilla {\it
et al.}, who studied samples with $\phi>0.58$
using DLS~\cite{brambilla09,elmasri09}.  Amazingly, they found that these slow and dense samples eventually equilibrated.  These results agreed
quite well with the earlier work of van Megen {\it et al.}  However, above $\phi_c$, the volume fraction where MCT predicts a
divergence of $\tau_\alpha$, they instead found finite values
of $\tau_\alpha$ -- suggesting that these samples were not
yet glasses and that $\phi_c$ of MCT is not equivalent to
$\phi_g$ for their samples.  These results are controversial
\cite{vanmegen10,brambilla10,reinhardt10,brambilla10b}; a key
issue seems to be the difficulty of determining volume fraction
and comparing results with differing polydispersity \cite{poon12}.

It should be noted that there are other predictions of how
$D_L(\phi)$ should behave near the colloidal glass transition.
A notable theory taking into account hydrodynamic interactions is
due to Tokuyama and Oppenheim \cite{tokuyama94,tokuyama95}, which
predicts a glass transition at a specific volume fraction $\phi_0
\approx 0.5718$ (with an exact expression given for this value).

Some understanding of how structure and dynamics relate to each
other has been recently presented by van Megen, Martinez, and Bryant
\cite{vanmegen09,vanmegen09b}.  They studied sterically stabilized
PMMA particles in decalin, which were well-characterized to behave
as hard spheres.  In \cite{vanmegen09}, they studied the mean square
displacement (MSD) and identified the time scale $\tau_m^{(s)}$
at which the MSD had the smallest logrithmic slope, that is, the
time scale at which the MSD was the most subdiffusive.  Here the
$(s)$ superscript indicates self-motion.  They also identified
an analogous time scale for collective motion, $\tau_m^{(c)(k)}$,
directly from DLS data, which was approximately the same as
$\tau_m$.  These time scales both grow as the glass transition is
approached, which makes sense.  By examining the $k$ dependence
of $\tau_m^{(c)}(k)$, they showed that structural arrest -- the
slowing of the motion -- starts at length scales corresponding to
$1/k_m$, where $k_m$ is the peak of the static structure factor,
and then spreads to other length scales \cite{vanmegen09}.  This
suggests that as the glass transition is approached, the spatial
modes of motion do not slow uniformly; that some are frozen
out sooner.  A related study by the same authors found that the
dynamics exhibit qualitative changes for the metastable states,
that is, $\phi > phi_{\rm freeze} = 0.494$ \cite{vanmegen09b}.

We refer the reader to~\cite{sciortino05,schweizer07} for good reviews
of experiments studying diffusion and relaxation times for the
colloidal glass transition, and how the experiments relate
to MCT predictions.  Earlier reviews summarize the state
of these questions in 1998 \cite{bartsch98cocis} and 2001
\cite{hartl01}.  Mode coupling theory is specifically reviewed
in~\cite{gotze92,reichman05}; see also~\cite{zaccarelli02}.

\subsection{Fragility}
\label{fragility}

\begin{figure}
\centerline{
\includegraphics[width=.8\textwidth]{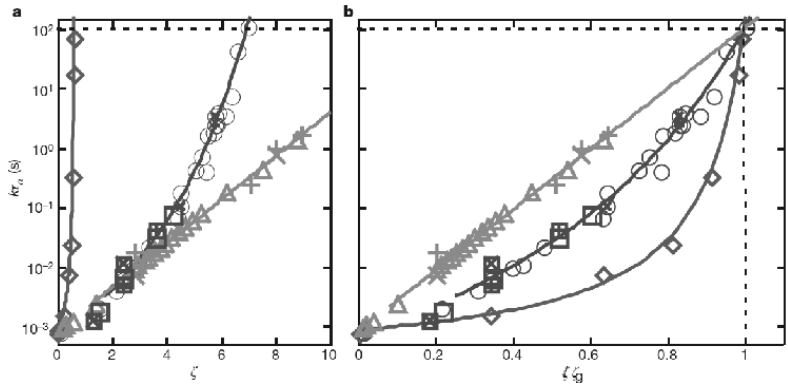}}
\caption{
(a) Plot of scaled relaxation time $k \tau_\alpha$, where
$\tau_\alpha$ is measured from light scattering, and $k$ is
chosen to collapse the data at low $\zeta$ values.  Symbols are
diamonds (stiff particles), circles and squares (intermediate stiffness), 
triangles (soft particles), crosses and pluses 
(rescaled shear viscosities from
rheology measurements, corresponding to intermediate and soft
particles, respectively).  (b) Same data as (a), with the
effective volume fraction $\zeta$ is normalized by $\zeta_g
\equiv \zeta(k \tau_\alpha=100$~s).
Reprinted by permission from Macmillan Publishers Ltd:
\href{http://www.nature.com/}{\it Nature}~\cite{mattsson09},
copyright 2009.
}
\label{fragilityfig}
\end{figure}

An important feature of molecular and polymer glasses is that, while the
relaxation time scale grows dramatically in all cases, the
{\it rate} of this growth varies between different samples.
This difference is termed the {\it fragility} \cite{angell95}.
Fragile glass-formers are ones in which the relaxation time scale
grows slowly over some range of the control parameter, and as the glass transition is approached, increases quite suddenly.  Such glasses are ``fragile'' in the
sense that when their viscosity is high, a slight change in the
control parameters (increasing $T$ or decreasing $P$) results in
a sharp decrease of the viscosity, easily ``breaking'' the glassy
behaviour.  In contrast, ``strong'' glasses exhibit Arrhenius
behaviour, where time scales and viscosity grow smoothly and
steadily as the glass transition is approached~\cite{angell00q}.
Results shown in figure~\ref{fragilityfig}(b) are from colloidal
glasses and illustrate the types of behaviours one might see:  here the
straight-line data are from a strong glass, and the curved data
correspond to fragile glasses.

The fragility can be defined in several ways.  One common way is to
fit the viscosity as a function of $T$ to the Vogel-Tammann-Fulcher
equation:
\begin{equation}
\eta/\eta_0 = \exp[D T_0 / (T-T_0)].
\end{equation}
Here $\eta_0$ is the viscosity at large $T$, and $T_0$ results from a fit
to where the viscosity would become infinite.  The parameter $D$ is called the ``fragility index'', and is larger for
stronger glasses; $D \rightarrow \infty$ corresponds to Arrhenius
behaviour. For fragile polymer glasses, $D$ can be as low as $\sim 2$~\cite{angell95}.  For colloids, one exchanges $T_0/T$ with $\phi/\phi_0$, given that for regular glasses
one decreases $T$ to $T_g$ and for colloids one increases $\phi$
to $\phi_g$.  Thus colloids would be fit with the Doolittle
equation with $C = 1$ [\ref{eqn:doolittle}]:
\begin{equation}\label{eqn:vft}
\eta/\eta_0 = \exp[D \phi / (\phi_0-\phi)].
\end{equation}
Using this definition, hard-sphere colloids appear
to be fragile glass formers with $D \approx 1.15$ (from
figure~\ref{chaikinviscosity} \cite{cheng02}) or $D \approx
1.65$ (from rheology \cite{marshall1990}).  Equivalent formulas
can be written using $\tau_\alpha/\tau_0$; light scattering
data on hard-sphere-like colloids suggest $D \approx 0.50$
\cite{brambilla09,elmasri09,kurita10b}.  An important question then is ``what features of a glass-forming material relate
to the fragility?''~\cite{kurita10b,hodge96,martinez01,mckenna02}.

This question has recently been explored by Mattsson
{\it et al.}~\cite{mattsson09} with colloidal suspensions of soft
hydrogel particles.  These particles easily deform and so can be
compressed as their concentration is increased; for this reason,
their glass transition does not occur at the same volume fraction of
$\phi_g \approx 0.58$ as for hard particles.  The authors considered
a generalized volume fraction $\zeta = n V_0$, where $n$ is the
particle concentration and $V_0$ is the volume of an undeformed particle.
Given that particles can be compressed to much less than
$V_0$, the generalized volume fraction $\zeta$ can greatly exceed 1.
Shown in figure~\ref{fragilityfig}, Mattsson {\it et al.}~found that softer particles (triangles) behaved as strong glasses, while harder particles (circles, diamonds) behaved as more fragile glasses.
This is an exciting demonstration of a model colloidal system which can be used to explore fragility.

It is not completely clear how particle softness relates to
fragility.  A simulation of different particle potentials did not
find any fragility changes \cite{michele04}, although this study
only considered varying $T$ rather than density.  Certainly for
polymer glasses, fragility can be quite different depending on if $T$ or
density is varied \cite{mckenna02}.  Furthermore, ``softness''
has two distinct meanings. In simulations such as
\cite{michele04}, softness
refers to the shape of the interparticle potential.  Often the
repulsive part of the potential decays as $1/r^n$, and smaller
values of $n$ are considered softer particles.  In contrast,
Mattsson {\it et al.}~used softness to refer to the particle modulus
\cite{mattsson09}, which probably is a prefactor to an
interparticle potential with fixed shape.

A recent mode-coupling study suggests that both the shape of the potentials and their prefactors may be needed to understand Mattsson {\it et al.}'s results~\cite{berthier10mct}.
For example, if the interparticle potential is $\sim 1/r^2$ at
large separations $r$, and $\sim 1/r^6$ at smaller separations,
then increasing particle concentration can shift the scale of
the interaction energy to a different regime of the potential.
If such an effect was properly accounted for, all of the data might
collapse for different softnesses \cite{xu09}.  In a sense, this
suggests that the results for the soft spheres can be considered
in terms of their effective hard-sphere size
\cite{barker76,berthier09epl}, thus
explaining the results in terms of a  mapping from
$\zeta$ to $\phi_{\rm effective}$~\cite{xu09,stieger04,eckert08}.  However, the functional form of
the interparticle potential is unknown for hydrogel particles.
Furthermore, it is not known if or how the interparticle
interactions vary between different batches of Mattsson {\it et
al.}'s particles.  (Other groups have noted that their particles
vary from batch to batch:  see discussion in \cite{eckert08}
comparing their results with their prior work in \cite{senff99}.)  
While the results of Mattsson {\it et al.}~are exciting, they raise many questions. A full understanding
requires either precise knowledge of how hydrogel
particles interact or insights from simulations on how to replicate the experimental data.

One can also consider the results of Mattsson {\it et al.}~in another way. The Arrhenius $\zeta$-dependence for the softer particles suggests
that the energy barrier for rearrangements is independent of
$\zeta$.  Perhaps soft particles can rearrange without significantly affecting others, and so rearrangements involve only a few particles.
In contrast the harder particles might exhibit growing dynamical
heterogeneity (see section~\ref{dynhet}), requiring more and more particles to rearrange,
and thus leading to a growing energy barrier with $\zeta$.  If
these conjectures are true, this would suggest that soft particles are not effectively hard particles with a different
radius.  Microscopy experiments may be able to shed light on the
question of particle rearrangements.

Recent simulations suggest different ways to tune the fragility
of colloidal systems.  One simulation showed that fragility was
tunable by using a binary system and controlling the size ratio
and number ratio of the two species \cite{kurita10b}.  While
intriguing and potentially useful for tuning the properties of
colloidal suspensions, a binary system would be of limited use for
understanding the fragility of single-component molecular glasses.
Another simulation studied soft particles with finite-range
potentials, quite analogous to soft colloidal particles,
and found that the temperature-dependent fragility increased
dramatically when the particles were over-compressed (density
increased above the point where the particles had to interact)
\cite{berthier09epl,berthier09}.

\subsection{Dynamical Heterogeneity}
\label{dynhet}

In a liquid below its melting point, some regions may exhibit
faster dynamics than others even though, spatially, these regions
may be very close \cite{ediger00,sillescu99,richert02}.  This
behaviour is called {\it dynamical heterogeneity}, and illustrates
that different regions of a system relax at different rates.
In such a system, relaxation time scales and length scales are
coupled, that is, longer relaxation times are typically associated
with larger collections of particles.  One key idea here is that of cooperative motions \cite{adam1965jcp}:  near the
glass transition, perhaps molecules need to ``cooperate'' in order
to rearrange.

\begin{figure}
\centerline{
\includegraphics[width=.75\textwidth]{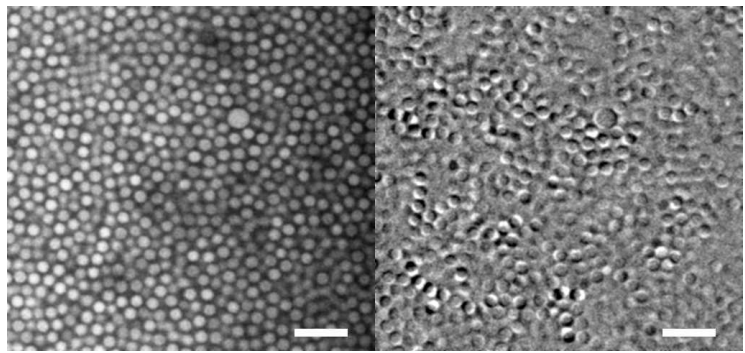}}
\caption{Left:  Confocal microscope image of a sample with
$\phi=0.46$.  Right:  Difference between the left image and one
taken 60~s later.  Where this image is gray, nothing has moved;
where it is black, a particle existed in the earlier time, and where it
is white, a particle existed in the later time.  Of course,
particle motion also occurs perpendicular to the plane of the image.
In both images, the scale bar indicates 10~$\mu$m.  For this
sample, the particles are slightly charged, shifting the onset of
freezing from $\phi_{\rm freeze}=0.494$ to $\approx 0.42$ \cite{hernandez09}.
}
\label{dynhetfig}
\end{figure}

Cooperative motion has been seen in colloidal samples, such as the one shown
in figure~\ref{dynhetfig}.  The left image shows a raw confocal
microscope image, a 2D slice through a 3D sample at $\phi = 0.46$.
The right image shows the difference between particle positions
in the left image and an image taken 60~s later.  Some regions of this image
are gray, indicating places where particles move relatively
little during this period of time.  Other regions are black and
white, indicating groups of particles all moving together.  For
these groups, motion is from black to white; for example, the
anomalously large particle moves slightly to
the left.  It can be seen that in general, neighbouring particles
that are rearranging tend to move in similar directions.  These
observations have been seen with microscopy in several colloidal
experiments
\cite{marcus99,konig05,kegel00,weeks00,mazoyer09,kaufman06,latka09}.

The interpretation of these results relate to cage trapping and
cage rearrangements.  As described in section~\ref{basicphysics}, at
short times, particles move due to Brownian motion, but this motion
is constrained because particles collide with their neighbours.
The neighbours thus ``cage'' the particle -- of course, the particle
is also part of the cage around its neighbours \cite{berne66,
sjogren80,wahnstrom82,gotze92,rabani97,verberg99,schweizer03}.
On longer time
scales, the cages relax and the system rearranges.  As figure
\ref{dynhetfig} shows, rearrangements often occur when one
particle moves, another particle follows, and so on.

\begin{figure}
\centerline{
\includegraphics[width=.55\textwidth]{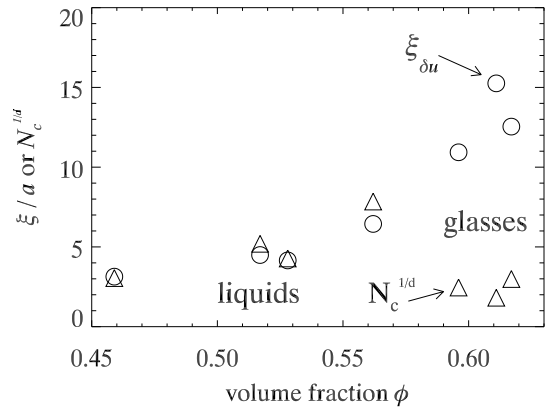}}
\caption{Data on growing length scales for dynamical
heterogeneity in colloidal samples.  Circles:  length scales
$\xi_{\delta u}$
found from spatial correlations of mobility, normalized by
particle radius $a=1.18$~$\mu$m~\cite{weeks07cor}.
Triangles:  mean cluster size in terms of number of particles $N_c$,
converted to a length scale using
the fractal dimension $d=1.9$~\cite{weeks00}.  These are clusters
of mobile particles as defined in~\cite{weeks00}, and by
computing $N_c^{1/d}$ we find the typical length scale of such
clusters.  The same data was analysed in~\cite{weeks00,weeks07cor}.
}
\label{corlength}
\end{figure}

As the glass transition is approached, the size of the cooperative
groups of particles increases, as well as the time scale for these
motions~\cite{weeks02sub,kegel00,weeks00,weeks02,weeks07cor}.
The growing length scales (quantified from spatial correlation
functions) extend up to $\sim 4-5$ particle diameters~\cite{doliwa00,weeks07cor}, shown by circles in figure~\ref{corlength}.
Rearrangements can involve up to $\sim 200$ particles~\cite{weeks00}, and the average size of a rearranging region is indicated by
triangles in figure~\ref{corlength}. Intriguingly, for liquid samples ($\phi < 0.58$),
these two length scales are essentially identical.  For glassy
samples, the correlation length scale is large, but cluster sizes
are small.  This is probably due to difficulty defining cluster
sizes in glassy samples as discussed in~\cite{courtland03}.
Rearranging particle displacements are small and can be lost
in the ``noise'' of particles diffusing within their cages in a
glassy sample.  More careful analysis reveals larger clusters in
glassy samples, although the details of defining such clusters
are ambiguous in glasses due to ageing \cite{courtland03} -- see
discussion in section~\ref{ageing}.

These experimental observations of spatial dynamical heterogeneity
are in good agreement with prior observations in computer
simulations \cite{doliwa98,kob97,poole98,donati99,doliwa00,
hurley95,hurley96,donati98,donati99pre}.  While the increasing
size of these cooperative regions is a striking observation,
it is a bit unclear exactly how this relates to the growing
relaxation time \cite{berthier04,karmakar2009} -- is it a cause,
effect, or side-effect?  Intuitively, it makes sense that if
more and more particles need to move simultaneously in some coordinated
fashion, that this is hard to do and will occur less often,
thus connecting directly to slowing time scales for diffusion:
in this sense, dynamical heterogeneity and the glass transition
are strongly connected, and the former could be said to cause the
latter.  Some evidence for this comes from
simulations which see diverging measures of dynamical heterogeneity
\cite{donati99,donati99pre,glotzer00b,lacevic03}.  One intriguing
recent result comes from simulations of a four-dimensional
hard-sphere system: there a glass transition was seen, but dynamical
heterogeneity was much less significant \cite{charbonneau10}.
This suggests that perhaps glassy behaviour can occur for other
reasons.  (As noted in section~\ref{simulations}, in general similar
phenomena are seen in 2D and 3D, and one would assume that a 4D
simulation can still provide useful insight into the 3D problem.)

It would be useful to understand the factors that allow some
particles to rearrange, or conversely the factors that prevent
the other particles from doing so.  It has been noted that
higher local volume fractions are weakly correlated with reduced
mobility~\cite{weeks02,conrad2005jcpb,kurita12}.  The correlation
is sufficiently weak that it has essentially no predictive
ability.  Simulations, though, have shed some light on this.
Widmer-Cooper, Harrowell, and Fynewever conducted simulations that
showed some regions within a sample have a higher ``propensity"
to rearrange~\cite{widmercooper04,widmercooper05}.  To come to
this conclusion, they ran repeated simulations beginning with
the same initial particle configurations.  From the initial
configuration, the system was evolved using molecular dynamics
and randomizing the velocities of the particles.  Though
again, these sites were only weakly correlated with structural
properties~\cite{widmercooper04,widmercooper05,matharoo06,berthier07}.
Such a procedure has not yet been tested in a colloidal experiment,
which would likely require using laser tweezers to establish a
known initial condition.  Investigating dynamical heterogeneity is
a good example where simulations are quite powerful; simulations
preceded and guided the experimental data analysis~\cite{donati98},
and allow for studies that are experimentally difficult or even
impossible~\cite{widmercooper04}.

Figure~\ref{dynhetfig} shows clear spatial variations in mobility.
However, another mode of studying dynamical heterogeneity
is to consider the temporal fluctuations of mobility:  in
any given region, the amount of motion will fluctuate in time
\cite{dinsmore01,courtland03,cipelletti03}.  These fluctuations
can be quantified with a four-point susceptibility function called
$\chi_4$ \cite{glotzer00b,lacevic03,berthier05,keys07,berthier07b}.
This function has been shown to be closely related
to the cooperative motion discussed above, and can
be used to pick out a time scale corresponding to
the dynamical heterogeneity.  The analysis has been
successfully applied to colloidal experiments several
times~\cite{berthier05,ballesta2008nat,sarangapani08b,narumi11},
with the results essentially in agreement with what has been seen
in simulations.

One could argue that studies of glassy systems should focus on
understanding the behaviour of slow particles, rather than fast
ones.  After all, over a given period of time the overwhelming
majority of particles in a glassy system are fairly immobile.
By studying mobile particles, one learns how mobility decreases as
the glass transition is approached -- for example, particle motions
are not only rarer, but also smaller~\cite{weeks02}.  Also, it is
usually the case in experiments that faster moving particles are
easier to distinguish.  However, one confocal microscopy study
focused its analysis on less mobile particles~\cite{conrad06},
finding that clusters of slow particles percolate through a
colloidal glass.


Dynamical heterogeneity in molecular glass experiments is reviewed
in~\cite{ediger00,sillescu99,richert02}.  Additionally, a chapter
in~\cite{cipelletti11book} gives a lengthy review of dynamical
heterogeneity in colloidal glasses.  An earlier review of dynamical
heterogeneity in soft glassy materials is~\cite{cipelletti02}.

\subsection{Confinement Effects}
\label{confine}

Phase transitions are usually investigated in the context
of macroscopically large systems.  However, confining
samples so that one or more dimensions are microscopic
leads to new physics, including confinement-driven phases
\cite{simionesco06}.  For amorphous phases, the glass transition
temperature $T_g$ is often changed by confinement
\cite{mckenna05,roth05,jackson91,barut98,richert94,morineau02,ngai02,torkelson03,torkelson05,roth07}.
In some experiments, the glass transition temperature is
decreased upon confinement (as compared with the transition
temperature in bulk) \cite{lowen99,kob00,richert94,morineau02},
whereas in others, the glass transition temperature increases
\cite{kob00,mckenna05,jackson91}.  In some cases, $T_g$ can
increase or decrease for the same material, depending on
the experiment \cite{kob00,mckenna05,morineau02,roth07}; this is likely due to differing boundary conditions~\cite{mckenna05}.
In molecular glass experiments, important differences are found
when studying confined samples supported by substrates, as compared with
free-standing films \cite{mckenna05,torkelson03,torkelson05,roth07}.
In other experiments, results depend on whether the confining surface is hydrophobic or hydrophilic. Computer simulations
indicate that confinement influences the arrangement of atoms
\cite{lowen99,kob00,yamamoto00,robbins92}, which might in
turn relate to the change of the glass transition temperature.
However, it is difficult to probe the structure and dynamics of
nano-confined materials.

Colloids thus can serve as an excellent model system for
studying confinement effects.  Such experiments have been performed by two
groups who confined samples between parallel glass plates
\cite{nugent07prl,sarangapani08}.  Nugent {\it et al.}
used a binary sample to prevent crystallization~\cite{nugent07prl}, while
Sarangapani and Zhu studied a monodisperse sample~\cite{sarangapani08}.  Both
experiments used confocal microscopy to observe a dramatic slowing
down of particle motion in samples that were very confined.  This suggests
that the glass plates act analogously to  ``sticky'' boundaries in the molecular glass experiments conducted on substrates, which
also find a slowing down of particle motion \cite{mckenna05}.
Follow-up work showed  that rough confining surfaces
slowed motions even further~\cite{edmond10b}.  The experiments
show a clear connection between layering of particles against the
walls and their mobility \cite{nugent07prl}, which has also been
studied by simulation \cite{lowen99,goel08,mittal08}.

\section{Features of Glassy Systems}\label{featuresofglasses}

\subsection{Amorphous Solids}\label{amorphous}

It is visually apparent from the bottom of figure
\ref{hardspherephasediagram} that colloidal glasses and crystals
have different structures.  Repeating patterns, like those in a
crystal, are completely absent in the glassy state and instead
the glass more closely resembles a very dense liquid.  Though
liquid-like, the system is dense enough that it can bear some degree
of stress over short time periods and respond elastically [see $G'$ in figure~\ref{masonfig}(a)].  Thus,
glassy systems are commonly described as {\it amorphous solids}.

\begin{figure}
\centerline{
\includegraphics*[scale=.6]{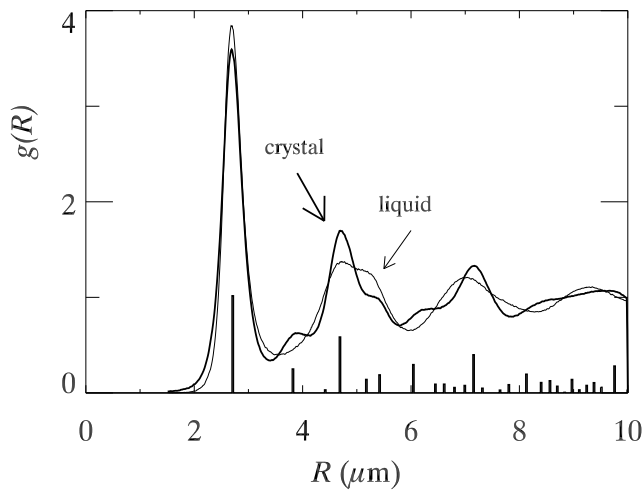}}
\caption{Pair correlation function $g(r)$ for a liquid sample
($\phi=0.48$) and a crystalline sample ($\phi=0.57$), with particle
radius $a=1.18$~$\mu$m.  The particles are the same colloidal PMMA
spheres used in~\cite{weeks00}, and were observed using
confocal microscopy.  For perfectly hard particles, the position
of the first peak in $g(r)$ would be at $2a$.  Here, however,
the position of the first peak is at $\sim 2.7$~$\mu$m, larger
than $2a=2.36$~$\mu$m, because the particles are slightly charged.
The vertical bars at the bottom edge of the graph are the
positions of the peaks of $g(r)$ for an ideal random hexagonal
close packed crystal.  Properly speaking, these should all be
Dirac delta functions (infinitely high); here they are truncated
by finite resolution, and rescaled but still proportional to
their magnitude in the ideal case.  The peaks of the experimental crystal are
broadened due to Brownian motion around the particles' lattice
sites, a slight polydispersity, and also particle tracking uncertainty.
}
\label{grs}
\end{figure}

A simple measure of structure is the pair correlation function
(or radial distribution function), $g(r)$, which describes
fluctuations in particle number density at a distance $r$ away
from a given particle.  Shown in figure \ref{grs} are the pair
correlation functions for a colloidal crystal and a colloidal
liquid.  The curve for the crystal (bold line) has fluctuations
at definite positions, corresponding to spacings between
particles in the random hexagonal close-packed lattice.  If this
were an ideal crystal, these fluctuations would be narrow spikes,
but in figure \ref{grs} they are broader due to Brownian motion, polydispersity,
and particle-tracking uncertainties, all of which ensure that
particle positions are not on exact lattice sites.  The liquid curve (thin line in figure \ref{grs}) differs significantly from the crystal, reflecting that the sample itself has much less
structure.  The second peak for the liquid, around 4~$\mu$m$ < r <
6$~$\mu$m, is slightly split into two sub-peaks.  As can be seen,
these correspond to two features of the crystal curve. The
origin of the sub-peaks is local packings of three or four
particles that appear crystalline and result in spacings of
second-nearest-neighbour particles that correspond to the
crystalline spacings.  In some experiments, this split second
peak of $g(r)$ is more obvious, and sometimes taken as a
signature of the glassy state; however, this is not a defining
feature but merely a common observation.

In 1991, Snook, van Megen, and Pusey used static light scattering to
study the structure of colloidal glass samples~\cite{snook91}.
Comparison with simulations of random close packed spheres confirmed
that the experimentally obtained colloidal glass was indeed
amorphous.  Later in 1995, van Blaaderen and Wiltzius used confocal
microscopy to examine the amorphous structure of a colloidal glass
\cite{vanblaaderen95}. They studied quantities such as
the number of nearest neighbours, bond-order parameters, and $g(r)$,
and found that all were in good agreement with simulations of
random close packing \cite{vanblaaderen95}.  More recently in 2010,
Kurita and Weeks used sedimentation to obtain a random close packed
sample and imaged nearly half a million particles (using confocal
microscopy), again finding that the experimental sample was quite
similar in many respects to simulations of random close packing
\cite{kurita10rcp}.  Their large sample size enabled a comparison to
recent simulations which found ``hyperuniformity,'' meaning that
density fluctuations disappear linearly with wavelength in the long
wavelength limit~\cite{donev05}.  An implication of a sample being hyperuniform is that it is incompressible, which has been conjectured to be a requirement for random close packed systems~\cite{torquato03}.  The only disagreement with simulations was that the experimental density fluctuations
did not go to zero at long wavelengths, implying that the sample was compressible.  However, recent simulation
results suggest that the disagreement is due to polydispersity in the
experimental sample~\cite{zachary11prl,zachary11pre1,berthier11}. Re-analysis of the Kurita and Weeks data suggests that accounting
for polydispersity does indeed demonstrate that the experimental
samples are hyperuniform, and thus incompressible~\cite{kurita11}.  
However, liquids and glasses have a finite compressibility:
these observations of incompressibility in random close packed
samples, then, demonstrate a structural difference between such
samples and liquids and glasses.  This shows the limits of such
samples as models for liquids, which is intriguing given that
random close packed hard spheres are one of the original models
for liquids \cite{bernal64}.

\subsection{Ageing}\label{ageing}

As with their molecular and polymer counterparts, colloidal
glasses exhibit \textit{ageing} -- as the system evolves
toward equilibrium, measured properties may change with time
\cite{mohan2010sm,hodge95,kob97aging,kob00aging,kob00epjb,vanmegen2008prl}.
In a sense, ageing can be thought of as a transient effect:  consider a
supercooled fluid whose temperature is decreased slightly from $T_1$
to $T_2$.  If the relaxation time scale at $T_2$ is $\tau_2$, 
then equilibration to the new temperature occurs over a period $\sim \tau_2$ \cite{vollmayrlee10}.  During the equilibration
process, the dynamics depend on the waiting time $t_w$ since the
temperature was changed to $T_2$.  These $t_w$ dependent dynamics are
ageing, and if $\tau_2$ is sufficiently large, then the sample can
be considered a glass which will age for as long as an experiment is performed \cite{warren2008epj}.  In contrast, for supercooled
liquids the same ageing phenomena are seen, but only for $t_w
\lesssim \tau_2$.

\begin{figure}
\centerline{
\includegraphics*[scale=.6]{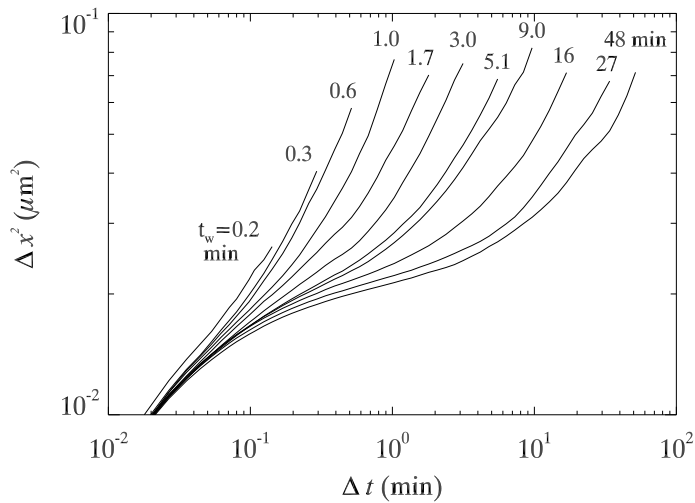}}
\caption{The mean square displacement for an ageing colloidal
sample.  The waiting time is labeled for each curve.  Data
from~\cite{courtland03}, corresponding to a sample
with $\phi \approx 0.62$ observed
with confocal microscopy.  Note that
each curve is averaged over a window of time centred on the
$t_w$ indicated; see~\cite{courtland03} for details of this
window averaging.  The curves are evenly spaced in $\log(t_w)$,
but not evenly spaced on the graph, reflecting the fact that
within the imaged region ($\sim 1000$ particles), ageing takes
place intermittently \cite{cipelletti03,castillo02}.
}
\label{agingmsd}
\end{figure}

Ageing is readily observed in colloids by examining the mean-square
displacement (MSD) at different waiting times, $t_w$, where the
waiting time is defined as the time since the last perturbation.
The system in figure~\ref{agingmsd} displays a slowing of dynamics
as $t_w$ is increased (with $t_w=0$ defined as the end of stirring)
\cite{courtland03}.  The short time dynamics are unchanged and
reflect particles diffusing within their cages. But with increasing
$t_w$, the plateau broadens and the upturn occurs at longer $\Delta
t$, indicating that relaxation occurs over increasingly larger time
scales.  Analysis of these data found that particle mobility was
related to $\log(t_w)$ \cite{boettcher11}.  Light scattering
data, in contrast, found time scales for motion which depended
algebraically on $t_w$ \cite{martinez08}.
For all of these experiments,
ageing occurs over the entire duration of the experiment, and so
these samples are easily classified as a glass.  There is recent
evidence, though, that even fairly dense colloidal samples can
eventually equilibrate given enough time, raising questions about
how to best define $\phi_g$ \cite{brambilla09,cipelletti2005jpcm}.

In molecular or polymer glasses, vitrification occurs when the
temperature is quickly lowered, viz. the system is thermally quenched.
In colloids, the creation of glassy systems typically involves
slow centrifugation of a sample.  Hence, the time at which
the system becomes glassy is poorly defined.  To this end,
the study of colloidal glasses often begins with a
process known as {\it shear rejuvenation} in which the system
is stirred or sheared in order to remove any history dependence~\cite{cloitre00,viasnoff02,bonn02,derec03,purnomo06}.  The hope
is that vigorous stirring breaks up any subtle structure in the
sample so that experiments begin with a randomized initial structure.  The time that the stirring stops
defines $t_w=0$ \cite{lynch08}.  Subsequent ageing, then, is presumably a slow evolution of the structure to some ``older''
state~\cite{cianci2006ssc}.  There is evidence that in some cases
the shear rejuvenation process produces different types of behaviour
than those observed in polymer glasses that are thermally quenched~\cite{mckenna03,purnomo06,mckenna2009rheo}.

A process more analogous to a thermal quench would be to increase
the volume fraction from $\phi_1$ to $\phi_2>\phi_1$.  This is
possible with hydrogel particles, whose size is controllable by
temperature (see section~\ref{softparticles}).  To date, there have
been several important ageing experiments using these particles
\cite{yunker09,di11}.  Purmono {\it et al.} studied the rheological
behaviour of ageing hydrogel samples \cite{purnomo06,purnomo07}.
They found that both shear rejuvenation and changing the volume
fraction resulted in reproducible initial states, although they
were slightly different.  Subsequent microscopy experiments by
the same group used tracer particles to confirm and extend these
results~\cite{vandenende10}; in particular, the sample was revealed
to be spatially dynamically heterogeneous, with mobile and immobile
particles coexisting.  Other experiments by this group demonstrated
that ageing behaviour was independent of particle softness, despite
the glass transition depending on both volume fraction and particle
softness~\cite{purnomo08}.  In another experiment, Yunker {\it et
al.} studied dynamical heterogeneity during the ageing of a quasi-2D
sample and found that the size of rearranging regions increased as
the sample grew older.  These results contrast with earlier work
done in 3D with a shear rejuvenated sample~\cite{courtland03};
differences in dimensionality and quench method are both plausible
explanations for the different observations.  In another hydrogel
experiment, Di {\it et al.}~\cite{di11} replicated certain classic
ageing experiments~\cite{kovacs64,zheng03}.  For example, in
1964 Kovacs found that molecular glasses quenched to $T_2$ from
$T_1>T_2$ would approach equilibrium differently  than if heated
to $T_2$ from $T_3 < T_2$.  Di {\it et al.} observed the same
`asymmetric approach' using hydrogel particles observed with DWS
(see section~\ref{dls}).

A natural question to ask is ``are there structural signatures
of ageing?'', that is, given information about the structure of
two samples, is there some quantity that distinguishes an ``old''
sample from a ``young'' one?  Indeed, it is somewhat intuitive
to expect a correlation between ageing dynamics and structure.
In polymer glasses, for example, systems become denser as they age
\cite{tant81}.  For a colloidal glass composed of hard-sphere-like
particles, presumably the only `clock' in the sample is the
structure; the sample should have no other way of knowing $t_w$.
The relevant quantity in colloidal glasses, however, has remained
elusive.  Cianci {\it et al.}~\cite{cianci2006ssc,cianci2006aip}
searched for correlations between the structure and dynamics of
an ageing colloidal glass, specifically looking at four-particle
tetrahedral configurations within the system (their sample was
quenched after shear rejuvenation).  A wide range of structural
parameters were examined but none changed with $t_w$.  However,
ageing did occur intermittently \cite{cipelletti03,buisson03} via
spatially localized groups of particles~\cite{courtland03,lynch08},
similar to what has been seen in simulations
\cite{kob00epjb,vollmayrlee02,vollmayrlee05,castillo07}.  These
slightly more mobile particles were associated  more disordered
local environments~\cite{cianci2006ssc}.  Presumably the more
mobile particles are the ones that change the structure and
thus are responsible for the sample having a larger age, but the
connection between structure, dynamics, and age remains unclear
\cite{cianci2006ssc}.  A similar set of observations was found
by the same group studying a binary sample: local composition,
quantified by relative numbers of small and large particles,
influenced dynamics but did not itself change as the sample aged
\cite{lynch08}.  These colloidal observations are in reasonable
agreement with simulations, which found that structural changes
during ageing are quite subtle~\cite{kob00aging,kob00epl}; they
also agree correlations between structure and mobility in other
simulations and experiments~\cite{weeks02,kawasaki07,royall08}.

Ageing has also been considered in regard to the fluctuation-dissipation theorem~\cite{cugliandolo97}, which connects
temperature, viscous friction, and inter-particle potential to
diffusive motion.  In the absence of inter-particle potentials,
as in the case of hard-sphere particles, fluctuation-dissipation
can be written simply as
\begin{equation}\label{eqn:flucdiss}
D = \frac{k_B T}{C},
\end{equation}
where $C$ is a viscous drag coefficient.  The drag force on
a sphere moving at velocity $v$ is $F_{\rm drag} = C v$ [see
(\ref{stokesdrag})].  Therefore, by measuring the velocity $v$
of a sphere feeling a known external force, and measuring $D$ for
the same sphere with no external force, one can calculate $T$.
In the same way that mean square displacement has a non-trivial
dependence on lag time $\Delta t$ and waiting time $t_w$
(as shown in figure~\ref{agingmsd}), so $D$ can be considered
to be a function of frequency ($\omega \approx 2 \pi / \Delta
t$) and $t_w$, and likewise the relationship between $F_{\rm
drag}$ and $v$ may have non-trivial frequency dependence
\cite{habdas04}.  The ratio of these quantities can be
generalized to provide an effective temperature, $T_{\rm eff}$
\cite{cugliandolo97}.  This idea has been tested in a wide variety
of nonequilibrium situations, including sedimentation experiments
\cite{segre01}, simulations of a sheared foam \cite{ono02},
granular experiments \cite{song05,wang08}, simulations of glasses
\cite{kob00epjb,kawasaki09}, experiments with regular glasses
\cite{grigera99,herisson02}, and experiments with colloidal glasses
\cite{buisson03,bellon02,bellon01,bonn03,wang06,bonn07}.  For hard-sphere colloidal glasses, the calculated effective temperature is a
few times greater than the actual temperature~\cite{bonn03,wang06}.
Though this may seem counter-intuitive, the observation can be
rationalized by arguments from statistical mechanics.  Temperature
is essentially a measurement of the width of the distribution of
energy states.  One can think of particle rearrangements as the
exploration of various energetic configurations available to the
system: for low to moderate $\phi$, rearrangements are easy and the
fluctuations in energy from one configuration to the next are small;
for $\phi \approx \phi_g$, rearrangements require cooperative motion
between many particles and therefore larger energy fluctuations.
As the energy distribution is broader for $\phi \approx \phi_g$,
this automatically implies a higher effective temperature.  Thus, the measured $T_{\rm eff}$ is related to cooperative
motion and large scale structural rearrangements within the
glass~\cite{wang06}.  Though the effective temperature is larger
than $T$, it was also found that $T_{\rm eff}$ does not seem to
change with age~\cite{wang06}.  

Recent experiments with magnetic particles~\cite{assoud09} allow
the possibility of doing experiments with an ultrafast effective
concentration quench.  These particles have a repulsion that
is controllable with an external magnetic field, and so their
effective size can be rapidly varied.  This enables experiments
to study ageing at extremely short time scales after the quench.


\subsection{Shear of Colloidal Glasses}
\label{shear}

So far, this review has discussed equilibrium properties of
supercooled colloidal fluids and out-of-equilibrium ageing in
colloidal glasses.  Another way to push a sample out  of equilibrium is to
apply shear, for example using a rheometer (see section
\ref{rheology}). The interplay of ageing and shear have been studied by
several groups \cite{koumakis08,fielding09,rogers08}, and
as discussed in section \ref{ageing}, shear is often used
by colloidal scientists to ``rejuvenate'' an ageing sample
\cite{courtland03,cloitre00,viasnoff02,bonn02}.  In this section, we will briefly summarize some
of the key results that have been found when shear is applied to
colloidal liquids and glasses.

One question to be asked is ``how quickly is the sample sheared?'',
which is quantified by the nondimensional P\`{e}clet number
\cite{chen10}.  This number can be thought of as the ratio of two
time scales.  The first is the time scale $\tau_D$ a particle
takes to diffuse its own radius $a$, defined in (\ref{taud}).
The second time scale is the inverse of the strain rate, $\tau_S =
(\dot{\gamma})^{-1}$, which is the time it takes for a particle to
be sheared a distance equal to its own radius, relative to another
particle in an adjacent shear layer.  Thus the P\`{e}clet number
is given by

\begin{equation}\label{peclet1}
{\rm Pe} = \displaystyle \frac{\tau_D}{\tau_S} =
\frac{\pi \dot{\gamma} \eta a^3}{k_B T},
\end{equation}

\noindent This assumes that diffusion is well described
by the physics in section \ref{basicphysics} [see
(\ref{msd3d})-(\ref{taud})].  More relevantly for colloidal
glasses, the long-time diffusion sets the time scale~\cite{habdas04,chen10}: in other words, the time a
particle takes to diffuse a distance $a$ is much slower than in a
dilute sample.  In this case, the modified P\`{e}clet number can
be defined in terms of the long-time diffusion constant $D_L$ as

\begin{equation}\label{peclet2}
{\rm Pe}^* = \frac{\dot{\gamma} \eta a^2}{6 D_L}
\end{equation}

\noindent [compare with (\ref{taud})].  If Pe$^* < 1$, then
diffusion is more important for rearranging particles, whereas
if Pe$^* > 1$, then shear rearranges particles before they can
diffuse an appreciable distance.  Thus, if one intends to understand
shear-induced behaviours, it is necessary to shear quickly (high
Pe$^*$) so that diffusion has insufficient time to equilibrate
the system.  In contrast, the measurements of $\eta(\phi)$ in
section~\ref{divergence} wish to avoid shear-induced effects
altogether, and so samples are sheared extremely slowly (low
Pe$^*$).

\begin{figure}
\centerline{
\includegraphics*[width=.7\textwidth]{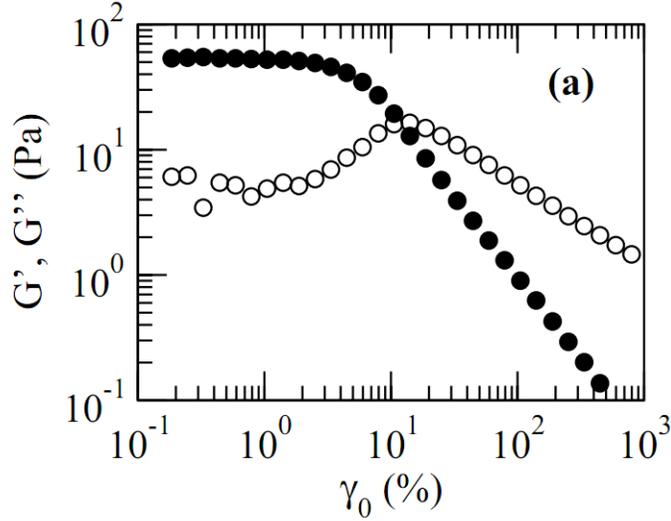}}
\caption{Elastic modulus ($G'$ - filled symbols) and loss modulus ($G''$ - open symbols) as a function of strain for a colloidal glass with $\phi = 0.645$.  The sample here has a polydispersity of $\approx 12\%$, hence the large $\phi$.  Squares correspond to a slow strain rate ($f = 1$~rad/s) and circles a high strain rate ($f = 100$~rad/s).  Figure reprinted from~\cite{petekidis03}. {\bf permission has been requested}.
}
\label{petestrain}
\end{figure}

The other question is how large a strain should be studied.
For small strains, the rheological behaviour is linear
\cite{mason95glass}; that is, the viscoelastic moduli do not
depend on the strain amplitude (see section \ref{rheology}).
Early measurements, such as those in figure~\ref{petestrain}, found that only above a critical strain $\gamma_c \approx$~15\%
does a colloidal glass sample show significant plastic rearrangements,~\cite{petekidis03,petekidis04} (a later experiment using confocal
microscopy lowered this to $\gamma_c \approx$~8\%~\cite{eisenmann10}).
This result supports the idea that a colloidal glass has a
yield stress, that is, it is a solid.  The elastic response
of the sample strained less than $\gamma_c$ has been attributed to a
distortion of the nearest-neighbour cages surrounding particles
(see section \ref{dynhet} for discussion of cages).
Equivalently, it can be thought of in terms of free energy;
in an unstrained sample, particles are packed randomly in some
structure that maximizes entropy and minimizes free energy.  When strained, particles move away from
that structure at a cost to free energy. So for small
strains, the sample gives a linear, elastic response.  Above $\gamma_c$, particles break free of their cages and can rearrange
\cite{pham06,chen10,petekidis03,petekidis04,brader10}.  (Although, it is
possible that a large region can shift without particles being
uncaged in the interior \cite{maloney08}.)  To some extent,
particle rearrangements occur for smaller strains, but may be
thermally activated in those cases \cite{schall07}.

Several works have noted that strained systems show
dynamically heterogeneity, similar to the thermally activated
rearrangements in supercooled colloidal liquids discussed
in section \ref{dynhet}.  In other words, the particles that
undergo plastic rearrangements in a sheared sample are distributed
nonuniformly in space. This has been studied in experiments
\cite{chen10,eisenmann10,schall07,petekidis02,besseling07,zausch08} and
simulations \cite{yamamoto98,shiba10,furukawa09,zausch08}, and was originally
predicted by theory \cite{falk98}.  One question that has been
studied is the shapes of these regions of rearranging particles:
are these rearranging regions oriented in any particular fashion
relative to the shear?  A sheared sample has three unique
directions: the velocity direction; velocity gradient direction;
and the vorticity direction (perpendicular to the first two).
Particle motions differ trivially along the velocity gradient
direction, and these overall affine motions can be subtracted from
the observed particle motion in a confocal microscopy
experiment \cite{besseling09,besseling07}.  To date,
confocal microscopy observations have found that diffusion is
isotropic \cite{chen10,besseling07} and that the shapes of the
rearranging regions are likewise isotropic \cite{chen10}.  However,
simulations suggest that particle motions and rearrangements are not
isotropic when sheared in the nonlinear regime~\cite{furukawa09}.
In general, there are a variety of ways to define and identify
plastic rearrangements in a sheared sample; these definitions are
compared and contrasted in \cite{chen10}.  The sizes of the
dynamically heterogeneous regions appear to be smaller in a sheared
sample as compared to the unsheared sample \cite{yamamoto98,zausch08}.

On a more macroscopic scale, a manifestation
of spatially heterogeneous motion is {\it
shear-banding}~\cite{manneville08,ovarlez09,schall10}.
Often when shearing complex fluids, the majority of
the strain occurs within a narrow band of the sample,
with the rest of the sample remaining relatively unstrained~\cite{manneville08,ovarlez09,schall10,dhont99,dhont03,dhont08}. Hence, a measurement of viscosity may only reflect only a small portion of a bulk sample.
Shear-banding has been observed and studied
in dense colloidal suspensions in recent years~\cite{fielding09,rogers08,chen10,moller08,besseling10,ballesta08}
and presented problems for rheology for much longer~\cite{buscall10}.  The origins of shear banding in complex fluids
continue to be an active research problem~\cite{divoux10}.

Given the industrial importance of shearing dense complex fluids, it
is unsurprising that the shear behaviour of many different types of
complex fluid glasses has been studied.  Examples include emulsions
\cite{hebraud97}, soft colloids \cite{legrand08}, star polymers
\cite{rogers10}, colloidal gels \cite{smith07}, and attractive
colloidal glasses \cite{pham08} (see section~\ref{otherglasses} for the distinction between a colloidal gel
and an attractive colloidal glass~\cite{zaccarelli09gel}).

There has also been a fair bit of theoretical work
studying the flow of glasses and glassy complex fluids
\cite{falk98,sollich97,fuchs07,crassous08,goyon08,bocquet09}.
Some of these are mode coupling theory analyses, which do a
good job describing shear of dense colloidal suspensions near
the glass transition \cite{fuchs07,crassous08}.  Other theoretical
work examines the flow of polydisperse glassy complex fluids
\cite{goyon08,bocquet09}.  A key prediction of this theory is
that a sample will initially build up stress elastically. After some time, a local
plastic event occurs and the stress is redistributed in a nonlocal
fashion around the location of the plastic rearrangement.
Thus the ability of a local region of the sample to flow depends
on what is happening in adjacent regions.  This may
potentially explain shear-banding \cite{bocquet09} and other
observations of flow profiles in a variety of complex fluids
\cite{goyon08,katgert10flow,nichol10}.

\section{Other Soft Glassy Materials}\label{otherglasses}

So far, this review is focused on the colloidal glass transition and experimental work on hard-sphere-like
colloidal particles, such as those used in the earliest experiments~\cite{pusey86,vanmegen86,vanmegen88}.  We
briefly here mention other soft materials that are used for glass
transition studies.

\subsection{Soft Colloids, Sticky Particles, Emulsions, and Foams}
\label{softparticles}

Several groups study colloidal glasses composed of polymer
hydrogel particles, which interact via a soft repulsion
\cite{yunker09,bartsch92,bartsch93,bartsch95,bartsch97,
bartsch02,eckert03,
mattsson09,di11,chen10nipa,eckert08,stieger04,siebenburger09,
dingenouts98,crassous08b}.  These particles can be packed to volume
fractions above $\phi_{rcp}$ of hard-spheres, with
potentially interesting consequences (see section~\ref{fragility}).  The stiffness of the particles can be controlled
by the amount of cross-linking or by using hard cores with
hydrogel shells.  The size of these particles can
be varied by controlling the temperature or pH~\cite{lyon00}.
Compressed particles are harder, while swollen particles are
softer:  this allows for the same particles to be used to test
how particle softness influences behaviour~\cite{mattsson09} and
to look for general trends \cite{siebenburger09,crassous08b}.
The temperature dependence provides a simple way to tune
the volume fraction {\it in situ}~\cite{yunker09,di11}.
This feature has already been exploited in ageing experiments
(see section~\ref{ageing}).  

One can also use sticky colloidal
particles to study colloidal gelation.  The most common way to
induce gelation is to add a small polymer molecule to the colloidal
suspension.  A colloidal particle surrounded by small polymers feels
an isotropic osmotic pressure whereas two particles close
together feel an unbalanced osmotic pressure that pushes them together. The result
is an effective attractive interaction between the particles, which
has been termed the depletion force~\cite{crocker99,asakura54}.
The range of this interaction is set by the size of the small
polymers (typically their radius of gyration is at least ten times
smaller than the colloidal particle radius), and the strength of
the interaction is set by the polymer concentration.

\begin{figure}
\centerline{
\includegraphics*[scale=.4]{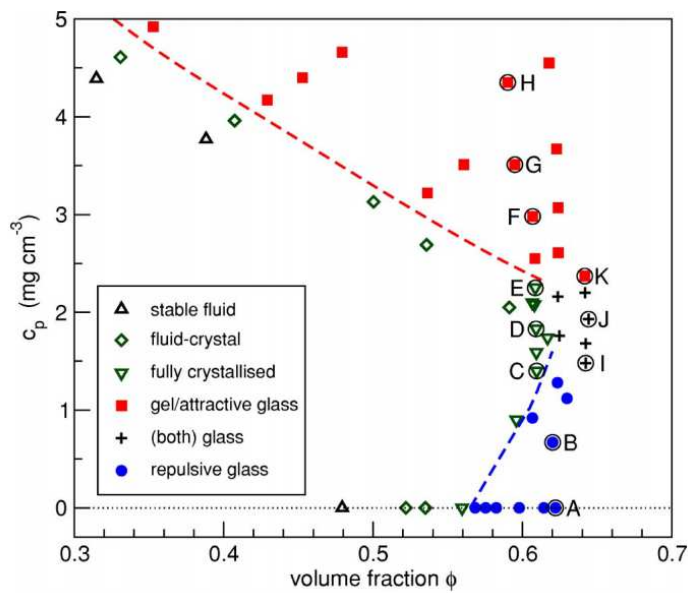}}
\caption{
(Colour online)
Phase diagram showing equilibrium and nonequilibrium states of a
colloid-polymer mixture.  Open symbols fully equilibrated,
while filled symbols did not and thus correspond to the glassy
states.  The dashed curves are guides to the eye.  
$c_p$ is the concentration of the added polymer, which had a size
ratio (polymer radius of gyration to particle size) of $\xi =
0.09$.  (The letters indicate specific states discussed in
\cite{pham04}.)
Figure reprinted from~\cite{pham04} -- {\bf permission has been requested.}.
}
\label{reentrant}
\end{figure}

Mode coupling theory was used to make intriguing predictions
of a re-entrant glass transition for hard-
spheres with a very short range depletion attraction
\cite{fabbian99,fabbian99err,bergenholtz99,dawson00,sciortino05},
with supporting evidence from simulations \cite{puertas02}.
A fascinating set of experiments confirmed these predictions
\cite{pham02,eckert02,bartsch02,eckert03,eckert04,pham04,pham08}.
A state diagram is shown in figure~\ref{reentrant}:  the filled
symbols indicate glassy states.
The re-entrant glass transition can be understood by considering the
two types of glasses.  At high polymer concentration, all of the
colloidal particles stick together in a gel -- an ``attractive
glass'' -- indicated by the filled squares in
figure~\ref{reentrant}.  Within the gel, particles are at a high local volume
fraction, whereas the pores of the gel are at a lower local volume
fraction~\cite{lu08}.  If the polymer concentration is decreased,
at some point, particles detach from the gelled phase and rearrange,
passing through the pores, and the sample becomes ergodic:  that
is, a fluid!  These states are shown by the inverted triangles in
figure~\ref{reentrant}, and eventually these samples all
crystallized.  If the polymer concentration is further decreased,
the distinction between the gel and the pores is lost, and the
system can become glassy again -- a ``repulsive glass'' --
indicated by the filled circles in figure~\ref{reentrant}.  At this
point, it is like the hard-sphere glass transition, where particle
arrest is due to cages.  Here the sample is much more isotropic,
with the local volume fraction essentially constant and larger
than $\phi_g$.  The intermediate fluid state allows for the possibility of high-volume fraction colloids with moderately low viscosities \cite{pham08,willenbacher11}.

For more about the physics of colloid-polymer mixtures, two
classic articles are \cite{lekkerkerker92,ilett95} and good
reviews are \cite{poon02,dawson02,sciortino05}.  Microscopy studies of
particle motion comparing attractive and repulsive glasses include
\cite{kaufman06,latka09}.  Note that for very strong attractive
forces and lower volume fractions, the sample is considered
a colloidal gel rather than an attractive colloidal glass; the
distinction is clarified in \cite{zaccarelli09gel}.

A more exotic colloid is laponite: nanometre-sized colloidal clay
platelets.  These aqueous suspensions have been observed to become viscous at very low
solid concentrations~\cite{mourchild95,mourchild98}  and exhibit
a glass transition at remarkably low volume fractions of $\phi
\approx 0.01-0.02$~\cite{bonn99,bonn01,joshi07,mohan2010sm}.
Along with rheology techniques, it is possible at such
low concentrations to use dynamic light scattering to study
dynamics.  One can also shear the suspension quickly to create highly reproducible
initial liquid states.  Given these advantages, laponite
suspensions are commonly used in studies of ageing, in particular
\cite{bonn99,bonn01,bonn02,buisson03,bellon02,bellon01,bonn07,munch03,leheny04}.
Many of these studies also relate to effective temperatures during
ageing; see section \ref{ageing} for more discussion of effective
temperatures.  

Emulsions are composed of droplets of one liquid immersed in
a second immiscible liquid.  The interfaces between the liquids are
stabilized by surfactants; review articles discussing emulsions
include~\cite{bibette99,mason06}.  Emulsions are like soft
colloidal particles in that the glass transition appears to be
closer to random close packing \cite{mason95emul,mason97emulsions}.
These have recently been used to study the forces between droplets
in close-packed samples \cite{brujic03,brujic03b,zhou06} and
the structure of random close packing of polydisperse samples
\cite{brujic07,clusel09}.

Foams, bubbles of gas in a liquid, are yet another system that
exhibits glassy behaviour~\cite{ohern01,katgert10}. A common
technique uses shear to study how the behaviour within the
foam depends on shear rate.  Shear in this case is used to
``unjam'' the foam, in contrast to increasing temperature
to liquefy a glass or decreasing the volume fraction to
unjam a colloidal glass.  This method has been used to good effect
\cite{durian95,saintjalmes99,langer00,debregeas01,lauridsen02,kabla03,dennin04,dennin08}.
One particularly thought-provoking simulation examined several
different measures of an effective temperature of a sheared
foam and found good agreement between the different measures
\cite{ono02}; see section \ref{ageing} of this review for further
discussion of effective temperatures.

Spin glasses, not yet mentioned in this review, are another large category of glass-forming systems~\cite{edwards75}.  There, glassy behaviour is controlled by range of interaction and the coupling strength between spins.  It has been demonstrated that confocal microscopy observations of colloidal glasses can be mapped onto spin glass formalism~\cite{chamon08}.

\subsection{Future Directions: Anisotropic Colloidal Particles}
\label{anisotropic}

\begin{figure}
\centerline{
\includegraphics*[scale=.4]{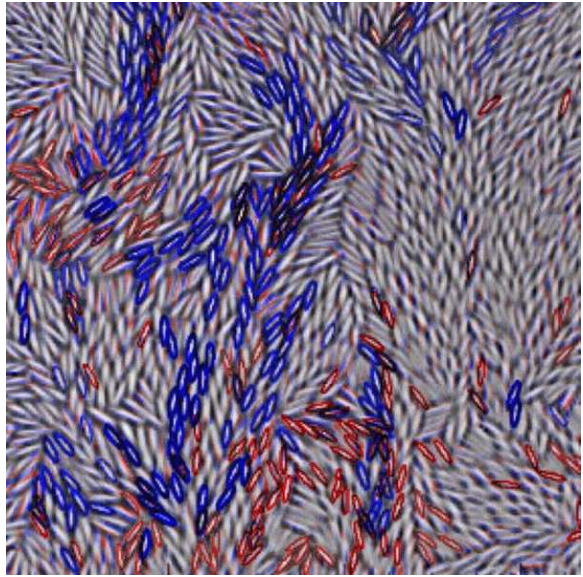}}
\caption{
(Colour online)
Ellipsoidal particles in a quasi-two-dimensional experiment, image
adapted from~\cite{zheng11}.  The ellipsoids that translate the
most at this moment are outlined in blue (thick darker gray lines),
and those that rotate the most are outlined in red (thin lighter
gray lines).  Those outlined in black translate and rotate.
The ellipsoids are 3.33 $\mu$m long.  Figure reprinted from~\cite{weeks11} -- {\bf permission has been requested}.
}
\label{ellipsoids}
\end{figure}

In recent years, a number of new methods have been
successful in synthesizing non-spherical colloidal particles
\cite{manoharan03,manoharan04,mohraz05,champion07}.
Several of these methods can produce
particles suitable for colloidal glass studies; see for
example~\cite{gerbode08,elsesser11}.  Anisotropic particles are
potentially better able to mimic the molecular glass transition,
which in many cases occurs with non-spherically symmetric
molecules.  A recent experiment studied the glass transition
of colloidal ellipsoids in 2D \cite{zheng11} and found that
rotational motions ceased -- became glassy -- at an area fraction
lower than that for the translational motion.  This confirmed
theoretical predictions \cite{schilling97,sciortino05}.  
An image adapted from their experimental data is shown in
figure~\ref{ellipsoids}, indicating that these particles rotate
and translate in spatially dynamically heterogeneous ways (see
section~\ref{dynhet}).  Large translational motions typically
occurred for particles within aligned domains, while the large
rotation motions were more prevalent between the domains.
This sort of experiment could easily test other predictions:
for example, simulations and theory predict a nonmonotonic
dependence on aspect ratio \cite{letz00,yatsenko07,demichele07}.
The differing glass transitions for rotational and translational
motions should disappear for aspect ratios closer to 1
\cite{schilling97,pfleiderer08}.  Aspect ratios closer to 1 might
be useful to study, given that molecular glasses only have a single
glass transition \cite{weeks11}.

Other possibilities including using ``patchy'' colloids, which
have attractive regions on their surfaces.  Simulations suggest
interesting gel phases can form \cite{bianchi06,zaccarelli07}.
A good review of experimental efforts to create patchy colloids
and recent simulation results is~\cite{bianchi11}.

A different take on breaking spherical symmetry is to use particles
which are optically asymmetric, but spherical in other respects.
Such particles have been created and their rotational motion
can be tracked \cite{anthony06}.  This is a fun technique,
but probably needs to be combined with other techniques to give
such particles some other non-spherical symmetry; otherwise, the
spheres are unlikely to dramatically change their behaviour near
the colloidal glass transition.  While hydrodynamic interactions
with nearby particles slow rotational motion, the regular colloidal
glass transition, which is an inhibition of translational motion,
would be unaffected.

The many other ways to make non-spherically symmetric particles are
comprehensively reviewed in \cite{glotzer07}.  One could imagine,
for example, tetrahedral particles with two sticky patches,
perhaps with the sticky patches having some specified position on
the tetrahedra.  Studying the interplay of shape, interaction,
and the colloidal glass transition is likely to be fruitful for
quite some time.

\section{Conclusion}\label{conclusion}

Colloidal suspensions are used in many ways to study the physics of
the glass transition, as has been described throughout this review.
As noted in the introduction, dense colloidal suspensions have
material properties of potential interest and relevance to a variety
of industrial processes, and so even without a connection to the
glass transition, dense colloids are worthy of study.  Aspects of
the colloidal glass transition remain to be understood, such as
the exact roles of gravity and polydispersity in suppressing
crystallization, correlations between structure and dynamics,
and the best way to define $\phi_g$, among others.  Given the
rich physics in these systems and recent progress in synthesis
techniques, the field is likely to continue flourishing for some
time.  In this review we have noted other reviews where topics
are discussed in more detail, and so we conclude by mentioning
prior reviews of the colloidal glass transition, several of which
have been cited in relevant locations throughout this current
review~\cite{schweizer07,hartl01,dawson02,sciortino05,pusey08}.

\section*{Acknowledgments}
We thank R. Besseling, A. J. Liu, D. Reichman,
and C. B. Roth for helpful discussions.  This work was supported
by a grant from the National Science Foundation (CHE-0910707).


\bibliography{combined}

\end{document}